\documentclass[a4paper,11pt]{article}
\usepackage{jheppub}
\usepackage{lineno}
\usepackage{textcomp}

\title{Optimal operating parameters for next-generation xenon gas time projection chambers}

\author[1]{K.~Mistry,}
\author[1]{Y.~Mei,}
\author[1]{D.R.~Nygren,}
\affiliation[1]{
Department of Physics, University of Texas at Arlington, Arlington, TX 76019, USA}
\emailAdd{krishan.mistry@uta.edu}

\abstract{The next-generation of neutrinoless double beta decay ($0\nu\beta\beta$) searches are targeting half-life sensitivities towards 10$^{27}$--10$^{28}$ years. Gaseous xenon time projection chamber (GXeTPC) detectors may be able to meet this challenge due to their excellent energy resolution and background rejection power through event visualization. This paper explores how the design choices of a next-generation GXeTPC time projection chamber can impact the overall performance of the experiment. We study the performance of systems using xenon enriched in the isotope $^{136}$Xe or natural xenon, focusing on scenarios that incorporate one tonne of $^{136}$Xe isotope. The detector size, copper shielding mass, energy resolution, density (using pressure at 293~K for convenience), and corresponding levels of diffusion are surveyed to evaluate the overall performance dependencies on these parameters. A detector optimized for using enriched xenon is preferred over natural, due primarily to a factor of 10 lower background rate driven by the large intrinsic backgrounds introduced by the copper shielding used in the detector. The performance of three types of gas TPC technologies was also explored based on different gas additives used to reduce diffusion to different levels. For all TPC technologies, we find background rates of a fraction of a count per tonne year in the region of interest are achievable. These performance levels are contingent on suitable energy resolution and event position placement in the drift direction being achieved for the specific detector technology. Performance for enriched xenon TPCs varies mildly with pressure in the range 5 to 25 bars, reaching background levels below 0.2 events/tonne-year.  Performance at one bar is worse by approximately a factor of four. When considerations for the construction of the detector in addition to the selection performance are included, there may be no clearly optimum pressure. }

\begin{document}

\maketitle
\flushbottom

\section{Introduction}
\label{sec:intro}
Neutrinoless double beta decay ($0\nu\beta\beta$) is a hypothetical process in which two neutrons in a nucleus beta decay simultaneously, emitting two electrons and two protons but no antineutrinos. This radioactive decay is only possible if the neutrino is its own antiparticle (a Majorana fermion). In the simplest scenario, two virtual neutrinos annihilate within the nucleus with practically all the decay energy given to the two outgoing beta electrons. The experimental signature is a narrow peak at the summed beta energy end-point (Q$_{\beta\beta}$) of the Standard Model two-neutrino double beta decay ($2\nu\beta\beta$) spectrum. The discovery of the Majorana nature of the neutrino would address some of the most important questions about the Universe, including why it is matter-dominated and why the neutrino mass is so light.

The best limits for $0\nu\beta\beta$ to date have been obtained by the KamLAND-Zen experiment using $^{136}$Xe \cite{jkf6-48j8} reaching 3.8$\times10^{26}$ yr half-life at 90\% confidence level (CL) and LEGEND-200 with ${^{76}}$Ge \cite{LEGEND200} reaching 1.9$\times10^{26}$ at 90\% CL. The next generation of $0\nu\beta\beta$ experiments aim to improve on these limits by a factor 10--100, targeting sensitivities of $10^{27}$--$10^{28}$ years, a feat which will require one to several tonnes of source isotope. 

Here, we consider the performance of a gaseous xenon time projection chamber (GXeTPC) in the search for $0\nu\beta\beta$ using $^{136}$Xe as the source isotope with a Q$_{\beta\beta}$ of 2.458 MeV. The NEXT collaboration employs a high-pressure GXeTPC with electroluminescent amplification to realize energy resolution at the 1\% full width at half maximum (FWHM) level. The latest iteration in this program, NEXT-100 \cite{NEXT100}, consists of 70.5 kg of $^{136}$Xe at 13.5 bar pressure, and is currently operating at the Laboratorio Subterráneo de Canfranc (LSC) with expected sensitivity of the order $10^{25}$ yr at 90\% CL. NEXT-100 incorporates  1 m diameter cathode and anode planes~\cite{NEXT:2023blw}, a high voltage system suitable for reaching 65 kV \cite{N100HVFT}, and a gas pressure system that can attain pressures up to 15 bar \cite{NEXT100}.

 Energetic charged particles of any type traversing a GXeTPC will ionize and excite atoms as they deposit energy. In the pure xenon used in NEXT, the excitations produce a prompt flash of scintillation light (S1) at wavelengths of $\sim$175~nm that is used to determine the time of interaction. An electric field ($E$) generated by applying a high voltage (HV) between the cathode and anode planes drifts the free ionization electrons towards the anode plane. Diffusion during drift degrades the quality of trajectory information. After reaching the anode plane, the ionization electron signal is amplified by electroluminescence (EL), producing a secondary, larger signal (S2).  The S2 signal is used for precise energy measurement and imaging of the diffused particle track. The combination of the timing and imaging information allows for a full 3D reconstruction of charged particle interactions depositing energy in the detector.

The goal of this work is to explore a variety of design and operation choices of a GXeTPC detector. We consider a detector with a fixed isotopic source mass of one tonne and a cylindrical geometry with length equal to diameter of size $L$. Key variables to be surveyed include pressures from 1--25 bar\footnote{We assume normal temperature (293 K) throughout this paper such that pressure is proportional to density.}, energy resolutions from 0.3\%--1.2\% FWHM, natural versus enriched xenon, and diffusion-reducing gas mixtures with the addition of molecular additives or helium. Detector sizes are determined by these parameters and vary substantially, from L = 2~m at 25 bar enriched xenon to L = 13~m for natural xenon at one bar.

The paper is structured as follows: Sec.~\ref{sec:Design} describes the design considerations for optimizing a tonne-scale GXeTPC, Sec.~\ref{sec:Detector} details the detector simulation and geometry, Sec.~\ref{sec:Analysis} describes the analysis. The results are shown in Sec.~\ref{sec:Enriched}, followed by a discussion and concluding remarks in Sec.~\ref{sec:conclusions}.

\section{Detector Design Considerations}\label{sec:Design}

In the conceptual design of the GXeTPC experiment,  major considerations include (i) design: trade-off balances affecting performance such as feasibility issues in mechanical and electrical construction of the detector, (ii) choices in detector technology: through the introduction of gas additives, gain mechanism and signal sensing, which can alter significantly how the TPC operates, and (iii) analysis challenges: the mitigation of the overall intrinsic background to negligible levels while maintaining sufficient signal efficiency. Among these challenges, one of the most important variables is the choice of gas pressure. Operation at 1 bar represents a unique use case, as there will be no net force exerted by the gas on the containment vessel.
 
\subsection{Detector Design Challenges}\label{sec:DesignChall}

\textbf{High-Pressure Gas Containment}: The vessel size scales non-linearly in cost and difficulty.  For example, going from  $\sim$100~kg mass at constant pressure to a tonne-scale detector of $\sim$1000~kg mass increases the linear detector radius by a modest factor, $\sqrt[3]{10}\simeq2.2$. However, the force exerted on the end-caps of the detector containment vessel scales with the square of the radius whereas the ring holding bolts scales linearly, making the flanges increasingly massive with pressure.  In addition, gas must be mainly confined within the active volume, complicating vessel design. 


\textbf{Detector Size}: Considering a cylindrical geometry fixed mass of source isotope, the cylinder length is given by $L = \sqrt[3]{4M/\pi\rho P}$, where $M$ is the total mass of $^{136}$Xe, $\rho$ is the density of xenon at normal temperature and pressure ($\rho = 5.987$ kg/m$^3$), and $P$ is the gas pressure. The detector length scales as $1/\sqrt[3]{P}$ , yielding smaller detector sizes with increasing pressure, favorable for a confined site. 

\textbf{Site Location}: For the $0\nu\beta\beta$ search, the experiment must be built deep underground to reduce the cosmogenic backgrounds. There are several candidate locations which could site the detector, for example: SNOLAB, Boulby, LNGS, SURF, and LSC~\cite{baudis2022}. The feasibility to construct the detector must account for the available facilities at these site locations. For example, SNOLAB and SURF can only be accessed through a mine shaft of fixed size, which sets a maximum size that any given part such as the end-caps or cylindrical vessel can be transported underground. A design for one bar is advantageous for these deep site locations as containment can be assembled underground from manageable low-mass components. However, higher-pressure is favored when considering any limited underground hall size. 

\textbf{Radioactive Burden}:   Gamma rays of $\sim$2.5 MeV or greater that enter the active volume can deposit energy close to the Q$_{\beta\beta}$ value and mimic signal events. A shield of low-background ($\sim 1$~\textmu Bq/kg) \cite{NEXT2021_SensitivityTonneNEXT} copper for the inner layer of the detector is most cost-effective way to reduce the rate to an acceptable level. About 10-12~cm copper is sufficient to attenuate 97-99\% of these backgrounds originating from outside the detector, such as the stainless steel pressure vessel or site materials. While this copper has low activity compared to materials such as stainless steel, the large mass needed results in it being the largest source of background for the experiment. Assuming a fixed copper thickness, the mass scales with $L^2\sim P^{-2/3}$, favoring higher-pressure.

Cosmic muons interact with the detector surroundings, leading to a source of neutrons that can be captured and activate materials. Neutron shielding can be provided through the use of water shields, which are likely easier to construct for smaller-sized detectors. 

\textbf{High Voltage}: For the detector with the same cylindrical geometry and isotopic mass, the pressure varies as $1/L^3$, while the maximum drift length scales as $L$. For example,  a drift field of $E/P=60$~V/cm/bar requires a HV of 36 kV for 1 bar, with 6~m drift but would be 150~kV at 10 bars and 2.5~m drift, favoring lower-pressure operation.

\textbf{Readout Plane}:  The readout plane needs to have a suitable light-collection efficiency to obtain the required energy resolution of $\sim1$\% FWHM. The number of readout sensors nominally increases with the square of the detector radius, favoring higher pressure.

\textbf{Track Clarity}: The track length scales as $1/P$, while the diffusion scales as $1/\sqrt{P}$. As the ratio of diffusion to track length is thus proportional to $\sqrt{P}$, this ratio favors lower pressure and improves the ability to separate signal from background due to the improved track clarity. 
 
\textbf{Isotope Safety}: Risk of catastrophic loss of xenon due to rupture of the vessel or gas system does not exist at 1 bar pressure. Higher pressures or operation with molecular additives require more complicated xenon recovery systems.  

\textbf{Detector Calibration}: MeV-scale gamma rays from high-energy radioactive sources such as $^{208}$Tl are likely to penetrate deeper into the active volume at lower pressure, while overall statistics will be reduced due to the lower cross section with density. Calibration with low-energy sources such as $^{83\textnormal{m}}$Kr will likely be harder to reconstruct at lower pressures due to the larger area of the detector and lower S1 reconstruction efficiency. 

\textbf{Additional Physics Probes}: In the event of positive observation of $0\nu\beta\beta$, this would prove the neutrino is a Majorana particle; however, the underlying physics that drives the decay would still be unknown. Such physics could be probed in GXeTPCs \cite{NEXT2025_VertexTagging} by identifying the decay vertex and extracting kinematic information such as the individual beta electron energies and angle. Extracting this information favors low-pressures where there is a higher track clarity.

\subsection{Detector Technology}\label{sec:DetTech}

The use of gas additives can significantly change the operation of the detector, impacting a variety of different properties,  introducing both advantages and disadvantages. We consider three modes of operation, considering gas additives that alter the diffusion extent but also the fundamental detector gas technology where certain technologies are more well-established than others. These are (i) noble element gas additives for an \textit{Electroluminescent TPC}, (ii) molecular gas additives for a \textit{Topology TPC}, and (iii) near diffusion-free tracks for an \textit{Ion TPC}.

\textbf{Electroluminescent TPC}: An electroluminescent TPC preserves the use of $\sim$ 175 nm VUV scintillation light and the use of the S1/S2 signals through the use of pure xenon or xenon-noble-element gas mixtures. This includes the well-established NEXT TPC technology~\cite{NEXT2021_SensitivityTonneNEXT}. Such detectors have demonstrated energy resolutions below 1\% FWHM and plausibly could reach resolutions towards 0.5\% FWHM~\cite{10.1093/ptep/ptaf066, Renner2019_NEXTWhiteCalibration}. Furthermore, the timing difference between the S1 and S2 signals is an effective method for obtaining 3D reconstruction of event positions that can be used for rejecting radon induced or other backgrounds to negligible levels (see Sec.~\ref{sec:analysisChall}).

The use of $^4$He can be used for reducing the transverse diffusion by about a factor of two compared with pure xenon while also being compatible with EL with reductions in the total yield by up to 2-3\% for 10-15\% helium by pressure~\cite{Fernandes2020_LowDiffusionXeHe,Felkai2018_HeXeMixtures, McDonald2019_XeHeDiffusion}. Cosmogenic backgrounds associated with neutrons may be mitigated by a further addition of up to 5\% $^3$He due to its high neutron capture cross section~\cite{Rogers2020_Xe137Backgrounds}.

\textbf{Topology TPC}: Molecular additives such as CO$_2$, CH$_4$, CF$_4$, triethylamine (TEA: N(CH$_2$CH$_3$)$_3$), or trimethylamine (TMA: N(CH$_3$)$_3$) can be introduced to the TPC and significantly reduce diffusion of electrons during drift.  Molecular degrees of freedom from vibrational and rotational states transfer electron energy gained from the drift field in inelastic collisions~\cite{Azevedo2016_XeDiffusion}. This reduction enables better spatial reconstruction of the charged particle tracks produced in the TPC and superior background rejection power at the analysis level compared with pure xenon. 

These molecular gas additives, however,  also significantly change the TPC operation, as they quench or absorb the VUV scintillation light and suppress EL. Furthermore, TMA can convert xenon excimers to ionization electrons through Penning transfers. This process enables higher avalanche gains and a reduction of the Fano factor~\cite{PhysRev.72.26}. 

The use of such additives for a tonne-scale gas detector will require significant further R\&D  to demonstrate suitable position reconstruction and sub-percent energy resolution with reduced or lost electroluminescence. For example, at modest percentages with CO$_2$ admixture, the EL can diminish to 70\% or 35\% compared with pure xenon for CO$_2$ levels 0.05\% and 0.1\%, respectively~\cite{Henriques2017_SecondaryScintillation,Henriques2019_ElectroluminescenceTPCs} with a reduction in the transverse diffusion by about a factor of three to four. At higher levels towards 5\%, both longitudinal and transverse diffusion would reduce by a factor three to ten compared with pure xenon, respectively. However, this must be balanced with the strong quenching of scintillation light. 

TMA, at a few percent admixture level, has one of the strongest diffusion-reducing effects with electron diffusion coefficients, $D_{L/T}^*$, as low as 0.3~mm/$\sqrt{\textnormal{cm}}\times \sqrt{\textnormal{bar}}$~\cite{Alvarez2014XeTMA}, a factor three lower in $D_{L}^*$ and a factor of 10 lower in $D_{T}^*$ compared with pure xenon. The TPC would be operated with proportional avalanche signal amplification, and exploiting its strong Penning effect and fluorescence at $\sim$280 nm \cite{Nygren2011IntrinsicResolution}. A useful estimate for the energy resolution with proportional avalanche gain can be obtained from an extrapolation of the resolution obtained with proportional counters filled with xenon and TEA (expected to behave similarly to TMA)~\cite{Ramsey1989}. In this work using a 22 keV $^{106}$Cd x-ray, an energy resolution of 8\% FWHM was obtained. Extrapolating to 2.5 MeV with a $1/\sqrt{\textnormal{energy}}$ dependence (a factor of over 111) a resolution $\sim$8/$\sqrt{111}$ = 0.76\% FWHM is obtained. This assumes each electron is multiplied independently of others and with statistically equivalent early fluctuations. Further work with TMA-xenon gas have reported extrapolated resolutions of 1.2\% FWHM at 1 bar with 30 keV x-ray source~\cite{Alvarez2014XeTMA}. Higher energy calibration sources with MeV energy in TMA-xenon gas mixtures at low pressure are yet to be explored, while at high pressures towards 10 bar, worsening energy resolutions reaching 3\% FWHM have been reported~\cite{pandaxIII,GonzalezDiaz2015XenonTMA}. 

Positional placement of events in 3-D may utilize the use of the diffusion spread of the energy deposits, which has a square root dependence with drift time. Alternatively, measurement of the positive ion signals that drift to the cathode~\cite{PaulLuke} can be used to determine the drift time difference relative to the ionization electrons. 

\textbf{Ion TPC}: Molecular additives can be introduced such that they capture ionization electrons or can undergo charge transfer with the ionized xenon~\cite{Nygren2007}. The resulting ions will drift to the cathode to be imaged with sub-mm diffusion over meter-scale drift lengths, allowing for almost diffusion-free tracking that would now be limited by the smallest pixelization achievable for the detector readout. Ion diffusion is not minimized at the same E/P as for electron diffusion, so optimization of the drift field should be considered in this mode and may be unfavorable. Similar to the Topology TPC operation, such detector technologies are a less well-explored domain, with potential losses to VUV scintillation light and detector fiducilization that must be considered. 

\subsection{Analysis Challenges}\label{sec:analysisChall}

The overall intrinsic rate of background is coupled with the choices made in the detector design challenges discussed in Sec.~\ref{sec:DesignChall}. The major backgrounds that can deposit energy near the Q$_{\beta\beta}$ include:

\begin{itemize}
    \item Standard Model (SM) Double Beta Decay ($2\nu\beta\beta$): This background produces two beta electrons where the summed energy of the betas is a continuous spectrum. This background is not considered in this work as it can be reduced to negligible levels with an expected Gaussian energy resolution less than 1\% FWHM. 
    \item $^{214}$Bi: The radioisotope $^{214}$Bi originates from the natural decay chain of $^{238}$U. This beta decays to $^{214}$Po that releases a number of deexcitation gammas to the ground state. One of these gammas has an energy of 2.448 MeV with branching fraction of 1.5\%. The largest source of this background comes from the inner copper shielding which has trace amounts of $^{238}$U assumed to be in secular equilibrium with $^{214}$Bi. The 3.2~MeV beta from the decay $^{214}$Bi (branching fraction 19.2\%) and the prompt (100~\textmu s) 7.8~MeV $\alpha$ from $^{214}$Po can also become a background if the decay is near the copper surface (referred to as \textit{Bi-Po}) and is not sensed as originating outside the fiducial volume. 
    \item $^{208}$Tl: The radioisotope $^{208}$Tl originates from the natural decay chain of $^{232}$Th with a branching fraction of 35.9\%. This beta decays to $^{208}$Pb that releases a number of deexcitation gammas to the ground state. One of these gammas has an energy of 2.615 MeV with branching fraction of 99.8\%. The largest source of this background also comes from the inner copper shielding which has trace amounts of $^{232}$Th assumed to be in secular equilibrium with $^{208}$Tl.
    \item Cosmogenic: The radioisotope $^{137}$Xe is produced from neutron capture on the source isotope $^{136}$Xe with neutrons originating predominantly from cosmogenic origin, such as muon spallation on the cavern rock. This isotope beta decays with a half-life of 3.8 minutes, producing a single electron with 4.17 MeV end-point energy in the gas volume. Neutrons can also activate materials such as the inner copper shielding, leading to a broad energy spectrum of gamma rays up to 8~MeV. These cosmogenic gamma backgrounds are prompt and can be suppressed to a sub-dominant level with a suitable muon rejection system with veto time $\sim$2 ms~\cite{NEW0DBD}.
    \item Radon: The radon isotope $^{222}$Rn originates from the $^{238}$U decay chain with a half-life of 3.8 days. The isotope can diffuse into the gas volume, decaying through multiple steps. About 95\% of these decays result in a charged state of $^{214}$Bi that drifts to and plates out on the cathode. This results in an additional source of Bi-Po background correlated with the activity of $^{222}$Rn. Reduction of this background is assumed to be subdominant, assuming accurate positioning of the event along the drift direction~\cite{NEXT2018_Radon}. The impact of this background without event placement is also investigated considering scenarios with a Topological/Ion TPC.
\end{itemize}

The requirement to reach sensitivities to the half-life beyond 10$^{27}$ years is to reduce these backgrounds to less than a fraction of a count per tonne of isotopic mass per year within the region of interest (ROI) energy window at the Q$_{\beta\beta}$ (count/tonne/year/ROI). 

This work separates the background rejection power as a function of three categories: (i) containment efficiency, (ii) energy resolution, and (iii) selection performance. 

The containment efficiency describes the fraction of events, signal or background, that deposit energy near the Q$_{\beta\beta}$ region and is predominantly affected by the choice of gas pressure and detector geometry. The optimal scenario is to have high signal containment and low background containment. We consider two scenarios: (i) a detector optimized for using xenon enriched in $^{136}$Xe to 90\% and (ii) a detector that uses natural xenon which has about 9\% $^{136}$Xe. A natural xenon detector may be employed to improve the signal containment efficiency and/or if obtaining enriched xenon is not feasible on the timescale of the detector construction. Furthermore, a natural xenon detector could be employed if initial operation is done at 1-tonne $^{136}$Xe mass and then the enrichment fraction is improved over time to include 5--10 tonnes of $^{136}$Xe.

The energy resolution of the experiment is strongly dependent on the method of signal amplification, such that the gain process is proportional to the initial number of ionization electrons. The most common method used in GXeTPC $0\nu\beta\beta$ experiments such as NEXT and AXEL is via electroluminescence which have achieved resolutions near Q$_{\beta\beta}$ in the range 0.7 -- 0.9\% FWHM \cite{10.1093/ptep/ptaf066, Renner2019_NEXTWhiteCalibration}.  

With the low-density of the xenon gas, MeV-scale electron tracks can travel about 30 cm at 10 bar ($\sim$ 300 cm at one bar). The selection performance using accurate topological reconstruction of these tracks is a powerful method for discriminating signal from background in GXeTPCs. The signature of signal events includes the identification of two electrons, while the backgrounds typically produce one electron tracks. These electrons scatter strongly within the gas as they lose energy. Near the end of the trajectory a larger amount of energy is typically deposited, This concentration is designated as a ``blob", which can be used to identify the number of electrons in a track.

\section{Detector Simulation and Geometry}\label{sec:Detector}

The detector simulation uses a simple cylindrical geometry (diameter equal length) implemented in the open-source NEXUS framework, a Geant4~\cite{ALLISON2016186, 1610988, AGOSTINELLI2003250} based simulation framework developed by the NEXT collaboration~\cite{NEXUS}. The cylinder has a volume of xenon gas with gas pressure varied from 1, 5, 10, 15 and 25 bar. This volume is surrounded with a 4 cm layer of copper to model the inner copper shielding. This thickness was chosen since the simulation at 1 bar was prohibitively slow, and is normalized to give the expected rate assuming 12 cm copper shield. 

\begin{figure}[hbt]
\centering
\includegraphics[width=\textwidth]{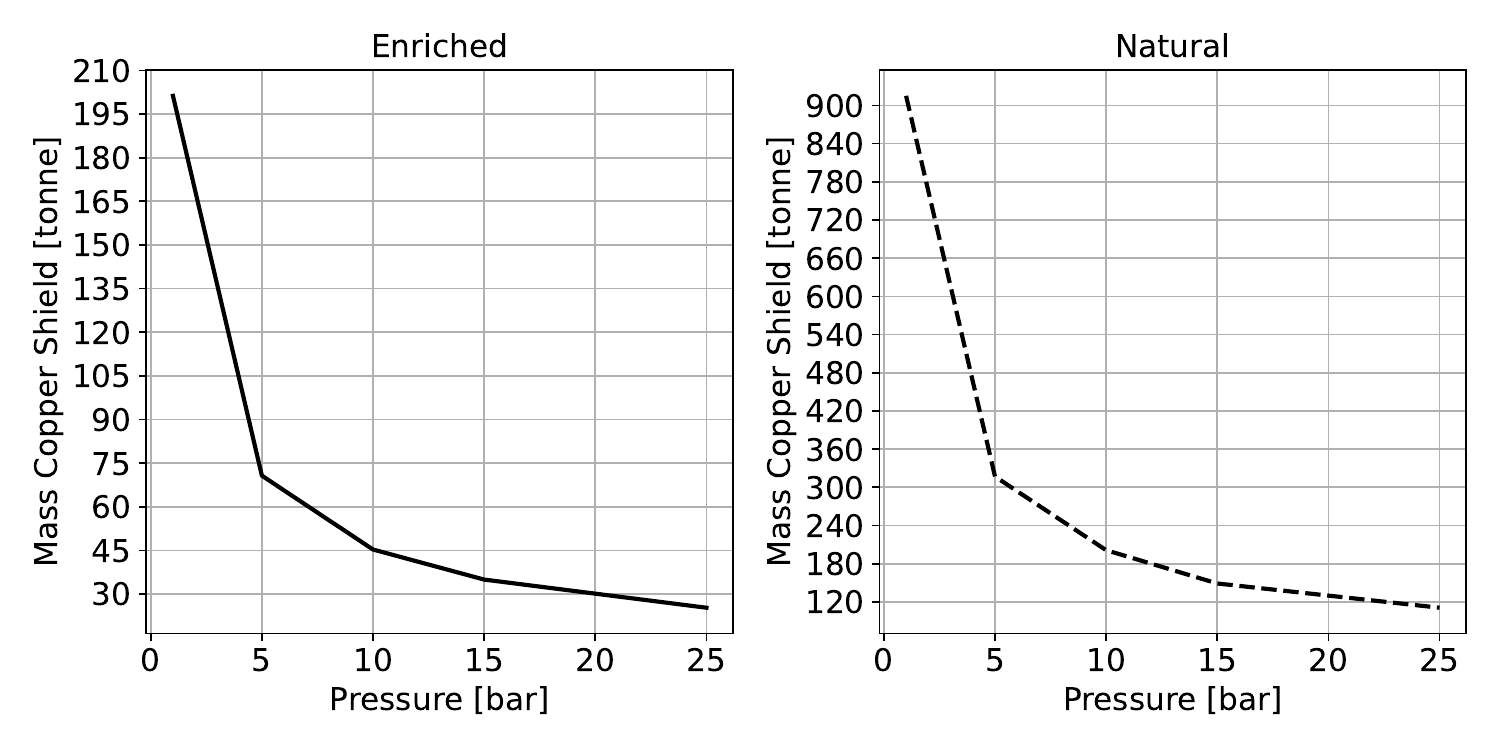}
\caption{\label{fig:masscu} The total mass of inner shielding copper as a function of pressure for (left) enriched and (right) natural detector assuming 12~cm thickness.    }
\end{figure}

The detector size is adjusted for changes in pressure to maintain 1 tonnes of $^{136}$Xe for either an enriched xenon detector (1-tonne module) or a natural xenon detector (10-tonne module). Table \ref{tab:pressure_params} details the simulated pressures and detector sizes, and Fig.~\ref{fig:masscu} shows the total mass of copper shielding required for 12~cm thickness. While these large dimensions appear unpromising relative to solid or liquid detectors of the same mass, the greater effective transparency of gas to gamma rays partly compensates.  

\begin{table}[h!]
\centering
\begin{tabular}{|c|c|c|}
\hline
\textbf{Pressure [bar]} & \textbf{Length (Enriched Xe) [m]} & \textbf{Length (Natural Xe) [m]} \\
\hline
1  & 6.2 & 13.3 \\
5  & 3.6 & 7.8  \\
10 & 2.9 & 6.2  \\
15 & 2.5 & 5.3  \\
25 & 2.1 & 4.6  \\
\hline
\end{tabular}%
\caption{Detector side-lengths at various pressures and enrichments. This assumes a cylindrical detector, with equal diameter to length and 1 tonne of $^{136}$Xe. The enriched assumes 90\% enrichment and natural assumes 9\% enrichment. }
\label{tab:pressure_params}
\end{table}

Events are generated in the gas producing energy depositions (``hits") along the track trajectory. For the $0\nu\beta\beta$ and $^{137}$Xe backgrounds, events are generated directly in the gas volume. For the gamma backgrounds, $^{214}$Bi decays and $^{208}$Tl decays are generated in the 4 cm of copper surrounding the detector. The number of interacting events is scaled down by a normalization factor of 0.37, constant with pressure, to give rates corresponding to 12~cm of copper. This factor was derived from generating 100 million $^{214}$Bi/$^{208}$Tl decays in the different thicknesses of copper.  

After generation, we save the information on the number of events depositing energy within a broad energy window of 2.3 -- 2.6 MeV. An energy smearing, assuming 1\% FWHM, is then applied to these events. In the next step, we save full event information, including energy deposits and particle information, only for events that deposit energy within a $\pm$1\% FWHM window of the Q$_{\beta\beta}$ value (2.43 - 2.48 MeV) in the gas region\footnote{Note, this $\pm$1\% FWHM window is equal to a total width of 4.71$\sigma$, such that 98\% of the signal events are retained. }. This pre-selection saves on computational resources for the next reconstruction stage, and the choice of 1\% window was to ensure there are sufficient statistics of events to analyze near the Q$_{\beta\beta}$. It is not expected that the topological signatures will vary appreciably for energy resolutions considered in this work ranging from 0.3--1.2\% FWHM. 

Different amounts of diffusion are then applied to smear the ionization electron positions based on a 3D Gaussian accounting for the longitudinal and transverse diffusion. The diffusion amounts are chosen to model plausible designs for an EL TPCs (xenon-helium), Topology TPC (xenon-CO$_2$), and Ion TPC (no diffusion). Across all pressures, 10\% He, 5\% CO$_2$, and no diffusion are simulated. The use of a larger variation of gas additive percentages for CO$_2$ allows for a slowly varying scan of different diffusions. Its application here can be extrapolated to other molecular additives, as described in Sec.~\ref{sec:DetTech}, with similar reductions in diffusion, such as TMA. This finer diffusion scan is studied at 1 bar only.

The diffusion amounts are summarized in Table \ref{tab:co2_diffusion_1bar} assuming a reduced drift electric field of 60 V/cm/bar.  This minimizes the diffusion in pure xenon and minimizes the longitudinal diffusion in xenon-helium gas mixtures. This drift field choice allows for a smooth scan of diffusion extent when varying the admixture of CO$_2$, although smaller drift fields may be preferable for  CO$_2$ admixtures. The cathode voltage requirements for this drift field are given in Tab.~\ref{tab:volatage_params}. 

\begin{table}[h!]
\centering
\begin{tabular}{|c|c|c|}
\hline
\textbf{Gas Mix} & \textbf{$D_L^*$ ($\sqrt{\text{bar}}$ mm/$\sqrt{\text{cm}}$)} & \textbf{$D_T^*$ ($\sqrt{\text{bar}}$ mm/$\sqrt{\text{cm}}$)}\\
\hline
Pure Xe  & 0.900 & 3.500 \\
10\% He  & 0.750 & 1.600 \\
0.10\% CO$_2$  & 1.307 & 0.818 \\
0.25\% CO$_2$  & 0.627 & 0.463 \\
5.00\% CO$_2$  & 0.314 & 0.300 \\
\hline
\end{tabular}
\caption{Longitudinal ($D_L^*$) and Transverse ($D_T^*$) diffusion coefficients as a function of CO$_2$ percentage/helium at a drift electric field of 60 V/cm/bar. Diffusion values are calculated from PyBoltz~\cite{ALATOUM2020107357} and with definitions in Ref.~\cite{McDonald2019_XeHeDiffusion}.}
\label{tab:co2_diffusion_1bar}
\end{table}

\begin{table}[h!]
\centering
\begin{tabular}{|c|c|c|}
\hline
\textbf{Pressure [bar]} & \textbf{Enriched Xe [kV]} & \textbf{Natural Xe [kV]} \\
\hline
1  & 37 & 80 \\
5  & 108 & 234  \\
10 & 174 & 372  \\
15 & 225 & 447  \\
25 & 315 & 690  \\
\hline
\end{tabular}%
\caption{The cathode voltage requirements for each detector type and pressure for 60 V/cm/bar drift field. }
\label{tab:volatage_params}
\end{table}

Given the large amount of ionization electrons ($\sim10^5$ produced for a 2.5 MeV track), a rebinning of the track into 3D voxels is applied with voxel sizes shown in Table~\ref{tab:voxelsizes}. Voxelization also models a finite pixelization of a realistic detector. The choice of voxel sizes in millimeters is estimated based on the average of the transverse/longitudinal diffusion amount, pressure dependence on diffusion, and maximum drift distance, such that the voxel size is roughly the size of a maximally diffused track:
\begin{equation}
\label{eqn:diff}
    \textnormal{voxel size} = \frac{D_L^*+D_T^*}{2}\times \sqrt{ L/P},
\end{equation}
where $L$ is the maximum drift length in cm, and $D_L^*/D_T^*$ are the longitudinal/transverse diffusion constants in [$\sqrt{\textnormal{bar}}$ mm/$\sqrt{\textnormal{cm}}$] as defined in Ref.~\cite{McDonald2019_XeHeDiffusion}\footnote{We note that another choice for the voxel size could account for the two transverse and one longitudinal degree of freedom: $(D_L^* + 2D_T^*) / 3$ or even voxels with different lengths in the drift and transverse directions.}. Since $L\sim P^{-1/3}$, Eqn.~\ref{eqn:diff} scales with $P^{-2/3}$. This also accounts for assuming a fixed $E/P$ such that the drift field scales in proportion with pressure.  The voxel sizes are rounded to the nearest integer except for the no diffusion sample. This sample uses non-integer bin sizes to try to preserve as much of the track detail as possible while keeping file sizes manageable. We do not explore the impact of varying the voxel size on the performance.

An energy threshold of 300 eV is also applied to remove the low-energy diffuse hits around track edges after smearing and reduce the file sizes to a more manageable size. The energy of these removed hits is proportionally redistributed to the rest of the track and involves typically less than 0.5\% of the total energy of the track.

\begin{table}[htb!]
\centering
\begin{tabular}{|c|c|c|c|c|c|c|}
\hline
\shortstack{\textbf{P [bar]} \\ \textbf{ }} & 
\shortstack{\textbf{10\% He [mm]} \\ \textbf{(EL TPC)}} & 
\shortstack{\textbf{5\% CO$_2$ [mm]} \\ \textbf{(Topology TPC)}} & \shortstack{\textbf{No Diffusion [mm]} \\ \textbf{(Ion TPC)}} \\
\hline
1  & 29 & 8  & 5   \\
5  & 10 & 3  & 2.2 \\
10 & 6  & 2  & 1.6 \\
15 & 5  & 1  & 1.3 \\
25 & 3  & 1  & 1   \\
\hline
\end{tabular}
\caption{The voxel sizes used in the analysis for all pressures. A voxel size of 55, 23, and 15 mm was used for the pure xenon, 0.1\% CO$_2$, and 0.25\% CO$_2$, respectively. These additional voxel sizes are studied for 1 bar pressure only.   }
\label{tab:voxelsizes}
\end{table}

\begin{figure}[hbt]
\centering
\includegraphics[width=\textwidth]{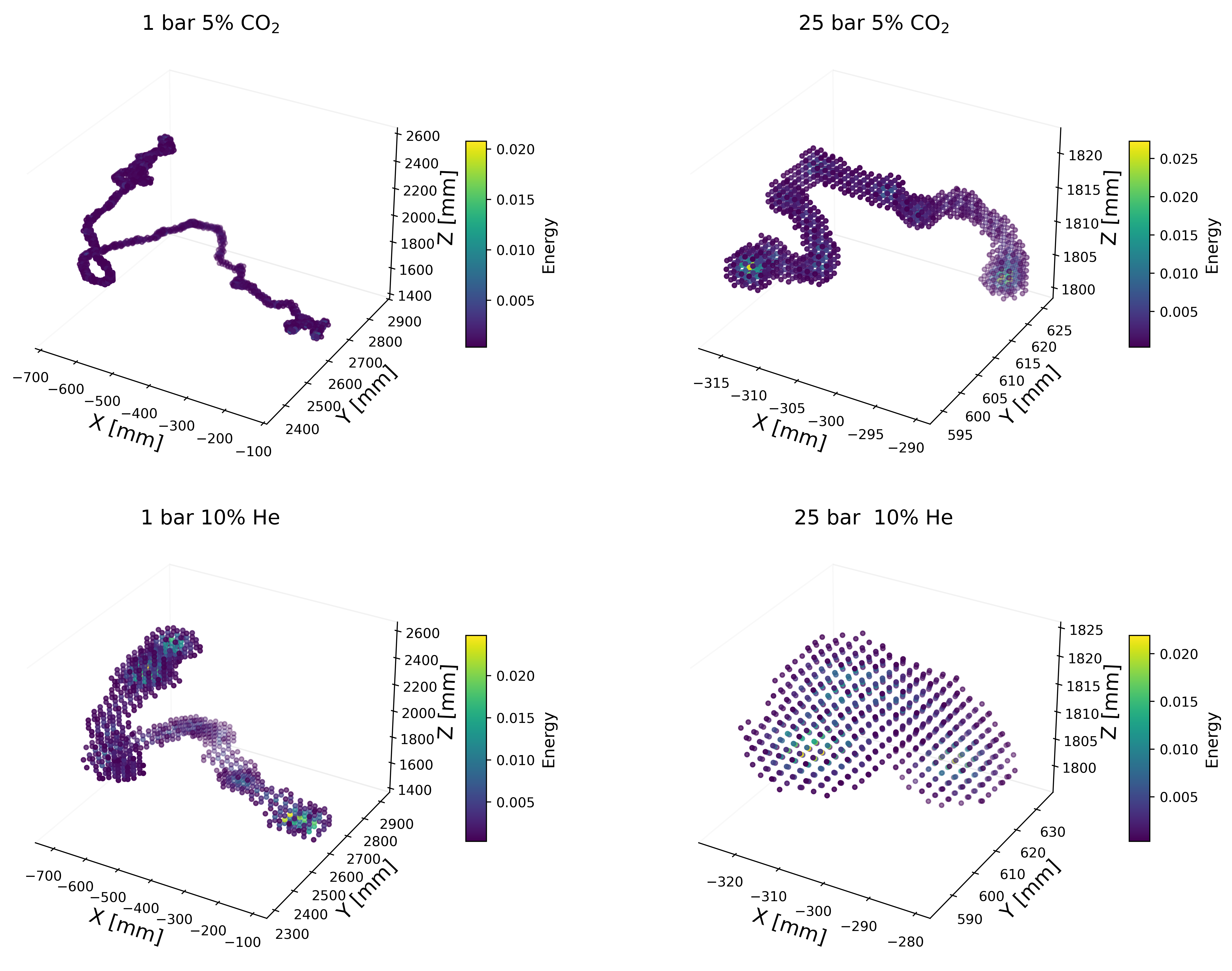}
\caption{\label{fig:DiffusionEvd} Example of the same signal event with different amounts of diffusion applied at 1 bar and 25 bar. These events include a 300 eV energy threshold, such that hits with energy below this threshold are removed and their energy is redistributed proportionally among the rest of the track hits. The voxelized event representation here indicates the track boundaries sharply. }
\end{figure}

The final event samples contain roughly 200k events for each background category within a window of $\pm$1\% FWHM of the signal peak (2.43--2.48 MeV). Examples of signal events with 5\% CO$_2$ and 10\% He are shown in Fig. \ref{fig:DiffusionEvd}. The visualization shows all the included hits, adding a coarse sharpness to track edges. 

\section{Analysis}\label{sec:Analysis}
As described in Sec.~\ref{sec:analysisChall}, the analysis stages can be broken down into three stages. For each stage, we define the acceptance factor, $\eta$, which is defined as the number of events selected divided by the total number of events in the initial or sub-sample. The total acceptance factor is then given by:
\begin{equation}
    \eta = \eta^{\mathrm{cont}} \times \eta^{\mathrm{Eres}} \times \eta^{\mathrm{sel}},
\end{equation}
where $\eta^{\mathrm{cont}}$ is the containment efficiency, $\eta^{\mathrm{Eres}}$ is the selection efficiency after applying an energy resolution cut, and $\eta^{\mathrm{sel}}$ is the selection efficiency using event topology cuts.

\subsection{Containment Efficiencies}\label{sec:Containment}

\begin{figure}[hbt]
\centering
\includegraphics[width=\textwidth]{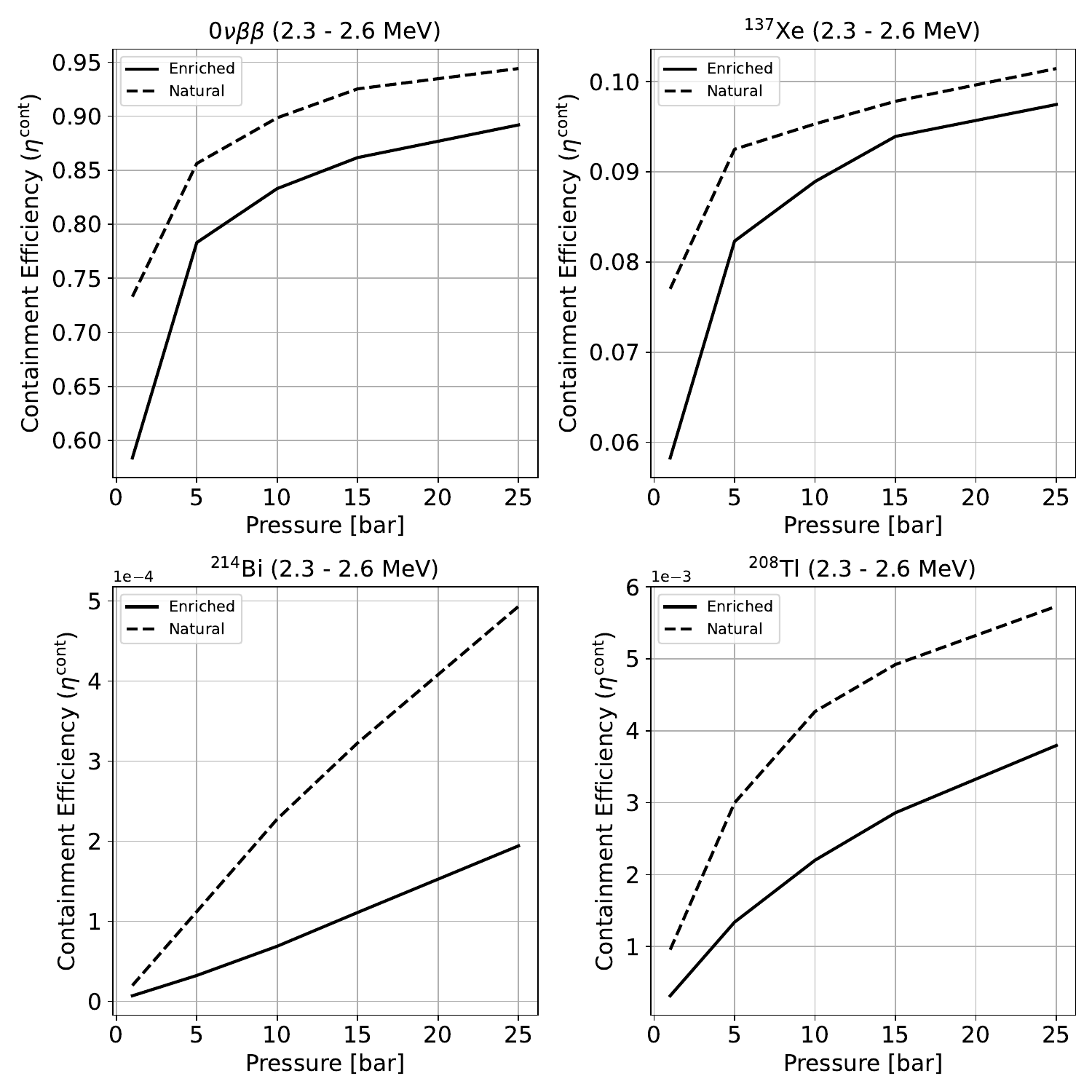}
\caption{\label{fig:Containment} The fraction of events depositing energy within 2.3--2.6 MeV to the number of decays of each radioisotope for an enriched and natural detector size.   }
\end{figure}

The event containment as a function of pressure for the signal and backgrounds and for an enriched and natural detector geometry is shown in Fig.~\ref{fig:Containment}. The containments are defined by the number events depositing 2.3--2.6 MeV in the gas volume divided by the total decays simulated. This range is chosen to allow a better understanding of the impact of energy resolution. The distributions are generated with a high-statistics simulation for each category, such that there are at least 5000 events depositing energy in the desired window. 

The signal events and $^{137}$Xe follow a similar trend. This is due to the similarity of the events of either one or two beta electrons being produced in the gas volume. The containment is a factor 10 lower for the $^{137}$Xe background compared with the signal because across the whole beta decay spectrum for this background, only about 9\% of betas will have an energy between 2.3--2.6 MeV.

The signal efficiency varies rapidly from 1 to 5 bar, with only 60\% of events contained at 1 bar, whilst it is 80-90\% for pressures above this. Lower signal containment efficiency reduces the strength that following analysis cuts can make due to the lower number of signal events. A natural xenon detector size has a larger volume, which increases the overall containment of the signal and background at lower pressure. 

At lower pressures, gamma backgrounds are reduced due to the product of much lower density and effective depth, leading to the lower containment fractions. However, interestingly, the $^{214}$Bi and $^{208}$Tl have a different shape with increasing pressure. The event distribution within the 2.3--2.6 MeV window can be seen in Fig.~\ref{fig:ERes}, top left. The majority of events from $^{214}$Bi consist of the photopeak of a 2.448 MeV gamma ray, while the majority of events from $^{208}$Tl consist of single and multiple Compton scatters of a 2.615 MeV gamma ray. 

The highest probability interaction mode of the $\sim$2.5 MeV gamma rays is via Compton scattering, occurring for about 80\% of all initial interactions. This initial scatter results in a second lower-energy gamma. For $^{214}$Bi events, this daughter gamma needs to deposit nearly all of its energy in the gas volume leading to the photopeak in the 2.3--2.6~MeV window. In the case of $^{208}$Tl, this gamma must re-interact but only deposit part of its energy in the gas volume as the energy window is below the photopeak of this gamma. At higher pressure, the partial energy deposition of the low-energy gamma is less likely compared to full energy deposition, and this effectively reduces the interaction rate at higher pressure. 

Overall, lower pressures are favorable for reducing the gamma backgrounds where 1 bar has almost a factor 10 and 30 lower containment compared to 25 bar for $^{208}$Tl and $^{214}$Bi, respectively. For all event categories, the containment fractions increase when going from an enriched xenon to a larger natural xenon detector size.

\subsection{Energy Resolution}\label{sec:ERes}

\begin{figure}[hbt]
\centering
\includegraphics[width=\textwidth]{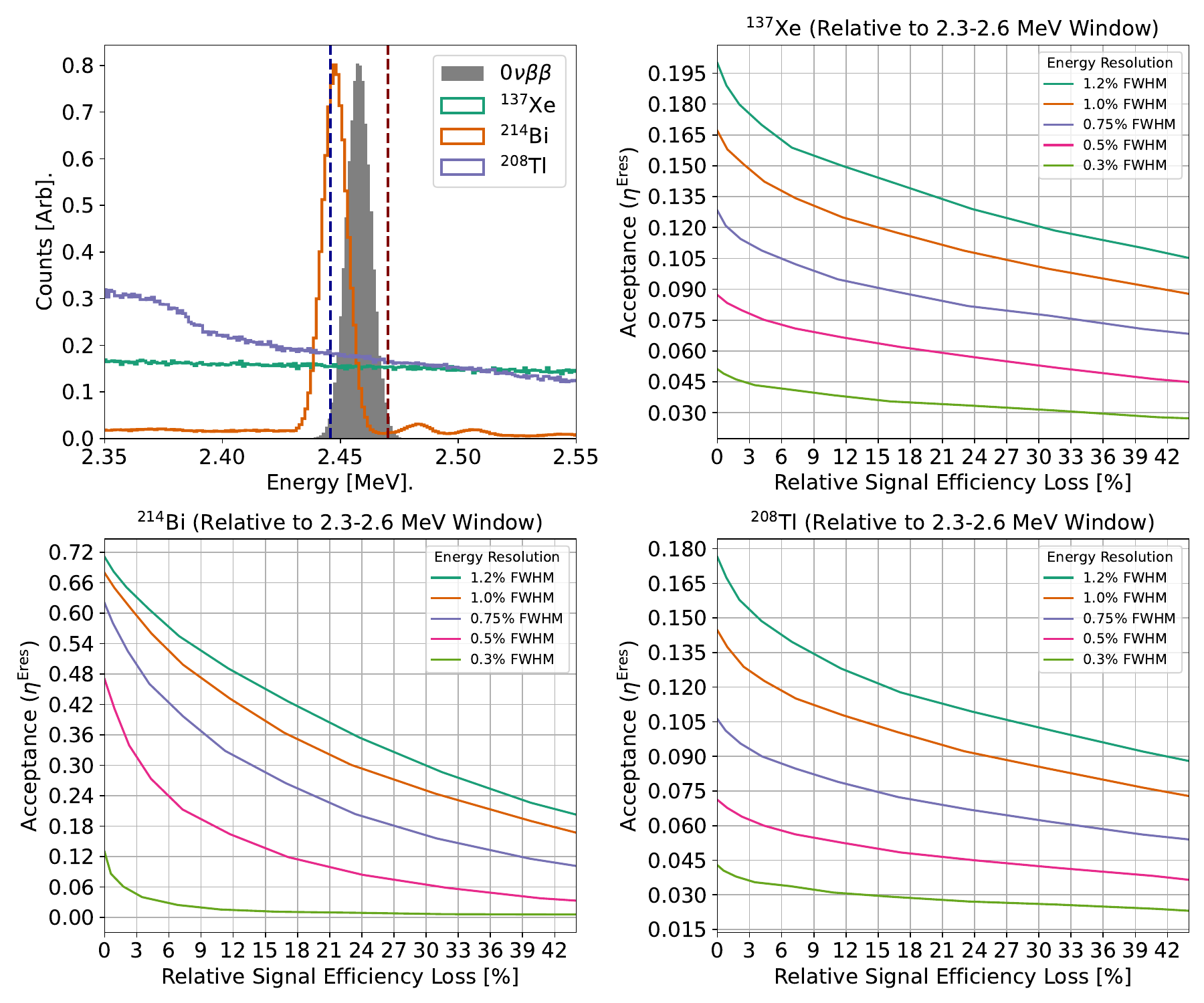}
\caption{\label{fig:ERes} Top left: The energy distributions of background assuming 0.5\% FWHM energy resolution. The blue and maroon dashed lines represent the bounds of one 0.5\% FWHM window. An asymmetric cut is done on this window by increasing the lower bound (blue line). Top right, bottom left, bottom right, show the background acceptance rates assuming different energy resolutions for $^{137}$Xe, $^{214}$Bi, $^{208}$Tl, respectively. The x-axis shows the efficiency loss from increasing the lower bound cut on energy.    }
\end{figure}

Excellent detector energy resolution is essential for reducing the $2\nu\beta\beta$ backgrounds to negligible levels, and is also powerful in reducing significantly other backgrounds. Fig.~\ref{fig:ERes}, top left, shows the distributions assuming a 0.5\% FWHM energy resolution. The dashed vertical lines mark around the signal peak shown in grey $\pm$0.5\% FWHM resolution corresponding to a ROI of 2.45--2.47 MeV.  

The $^{137}$Xe backgrounds are flat across this energy range, while the $^{208}$Tl has a slowly decreasing slope. The bump at $\sim$2.37 MeV corresponds to the Compton edge of the 2.615~MeV gamma ray. The $^{214}$Bi background includes the photopeak of the 2.448 MeV gamma along with a few higher energy $^{214}$Po deexcitation gammas. 

A poorer energy resolution increases the width of Gaussian smear applied to these distributions allowing more events to overlap with the signal. The overall acceptance of background is shown in Fig.~\ref{fig:ERes}, top right and bottom row for each background for different energy resolutions assumed for events depositing an initial energy within 2.3--2.6 MeV. The distributions are plotted as a function of a relative signal efficiency loss obtained from applying an asymmetric cut in the ROI. This cut is done by increasing the value of the lower bound energy range in the ROI window. This is better visualized by considering Fig.~\ref{fig:ERes} top left, whereby the dashed blue line is moved to the right. As this cut value is increased, this causes the total signal efficiency to diminish while also rejecting more background. 

The 0.3\% FWHM energy resolution represents the intrinsic resolution of a GXeTPC from Fano fluctuations \cite{PhysRev.72.26} assuming energy is measured from the ionization charge, while 1.2\% FWHM energy resolution represents an energy resolution an actual tonne-scale detector might easily achieve. More optimistic energy resolutions are in the range 0.5\% to 1\% FWHM. These distributions were generated with a dedicated simulation assuming 25 bar pressure; however, the shape of these distributions is  expected to vary insignificantly with pressure. 

The shape of the energy resolution curves are similar for $^{137}$Xe and $^{208}$Tl backgrounds as their shape is mostly flat while there is a stronger slope with $^{214}$Bi due to the variation of the energy cut along the photopeak.

\subsection{Reconstruction}\label{sec:Reconstruction}

Following the containment and energy resolution, the last stage in the analysis is to apply a selection to utilize the topology of the tracks. 

An initial fiducial volume is applied to all events, such that if any hit falls within 2~cm of the detector wall then it is rejected. This is particularly effective in removing the Bi-Po backgrounds. 

The reconstruction algorithm then builds tracks out of the energy-weighted center of diffused hits that are within a given proximity. The track consists of a set of contiguous points falling in a line. Some interactions may have multiple separated tracks, for example, when a gamma ray undergoes multiple Compton scatters or a $0\nu\beta\beta$ beta electron emits bremsstrahlung radiation that then deposits energy away from the main track. 

The primary track is defined as the track with the largest energy. In some cases, the electron may produce high energy delta electrons forking the track into three or more branches. In these cases, the primary track and information associated with this is defined as the longest segment from two end-points on the track. 

We extract several variables from the primary track, including its total energy and length. One of the most effective variables is the ``blob energy" which integrates the total energy in a sphere of radius $r$ around the track start and end. Blob 1 is defined as the blob with the highest energy, with blob 2 being the lowest energy blob. This variable is a key marker in identifying an electron stopping in the gas. Applying a minimum threshold to the blob 2 energy can remove a significant number of the gamma and $^{137}$Xe backgrounds as the track start of these single electron-like backgrounds does not generally produce a distinct blob.

The primary track energy and length variables are effective at removing events with significant bremsstrahlung and also multi-site interactions, as this reduces the overall primary track energy. It is effective to employ a single track cut to remove these multi-site events; however, this work opts to use the primary track information. This is because at 1 bar the single track cut leads to a 54\% signal efficiency loss when applied to the no diffusion and 5\% CO$_2$ samples and this results in a final efficiency below 25\%. This is due to a higher efficiency of tagging separated deposits near the track from bremsstrahlung. A separate analysis optimization was done for 5-25 bar employing a single track cut, and no significant difference in the overall performance was found using either method. 

The analysis is most efficient in selecting signal events when the energy difference between the two emitted electrons is smaller than 60-70\%. Furthermore, selection in drift distance is mostly flat for the 5\% CO$_2$ sample, while it exhibits a higher selection bias towards low drift distance for the 10\% helium sample where track clarity is highest.

The overall selection process is optimized such that the final signal efficiency is required to be about 25\% for all pressures and diffusions considered, so that they may be consistently compared. This final efficiency includes the containment efficiency and an assumed energy resolution of 0.75\% with an asymmetric cut with 18\% relative efficiency loss. 
At this efficiency, a 1-tonne $^{136}$Xe detector would yield approximately 1 signal count/yr while allowing for a suitable reduction in the background acceptance. This choice allows us to maintain a consistent comparison with varying gas density; however, other choices of final efficiency are also suitable depending on the desired signal to background ratio.

\subsection{Selection Diffusion Dependence}

\begin{figure}[hbt]
\centering
\includegraphics[width=\textwidth]{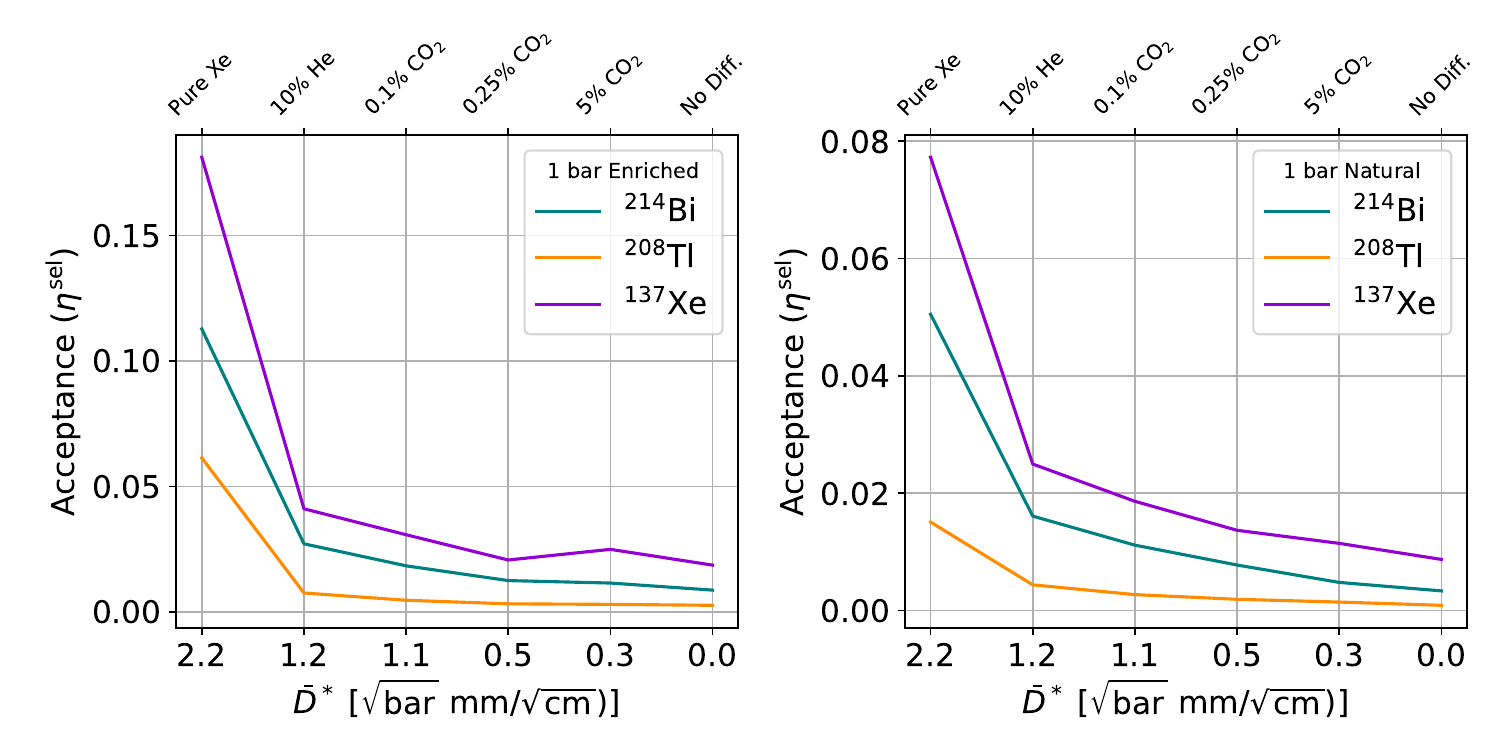}
\caption{\label{fig:SelAccDiff} Background acceptance factors for voxelized tracks at 1 bar for different diffusion amounts for an enriched (left) and natural (right) detector configuration. Overall, the dependence on diffusion has a similar functional form.  }
\end{figure}

To investigate the selection performance with diffusion, we consider 1 bar pressure scanning over a range of diffusion values listed in Table~\ref{tab:co2_diffusion_1bar}. The selection acceptance factors are shown as a function of a mean diffusion coefficient $\bar{D^*} = (D_L^*+D_T^*)/2$ in Fig.~\ref{fig:SelAccDiff}, considering an enriched and natural detector configuration. 

For an enriched detector, about a factor of three to four reduction in background rejection power is gained with gas mixtures that reduce diffusion relative to using pure xenon. The improvement with decreasing diffusion is expected since the overall track clarity and reconstruction of track information are improved. 

A natural detector has a similar trend to enriched and offers about a factor of two improvement in the overall selection power at 1 bar. This is primarily due to the higher signal containment efficiency, where stronger analysis cuts can be applied. 

\subsection{Selection Pressure Dependence}

\begin{figure}[hbt]
\centering
\includegraphics[width=\textwidth]{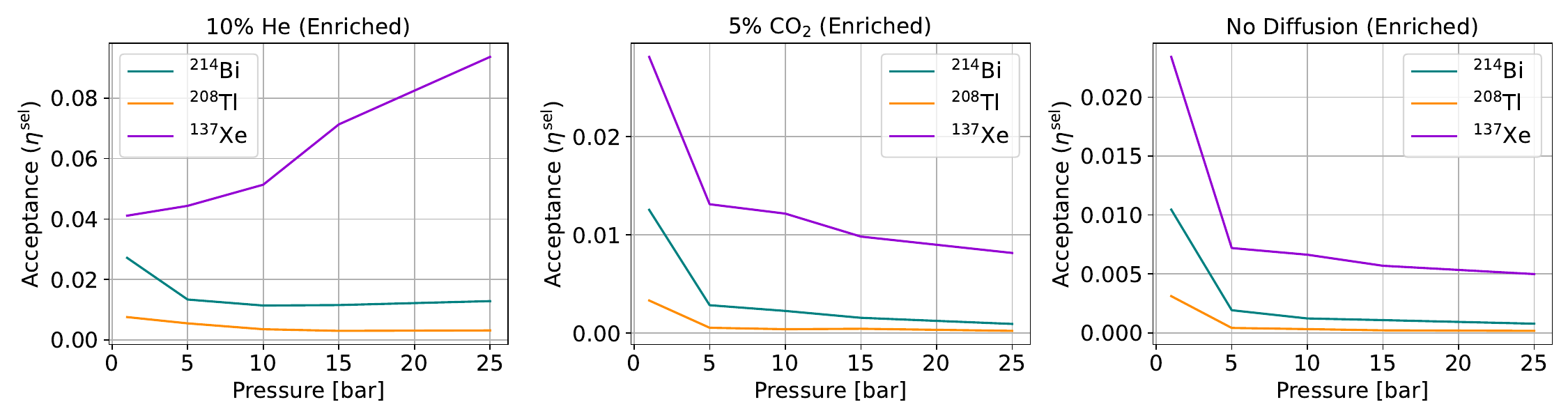}
\caption{\label{fig:EnrSelPerfPress} The acceptance factors from the topological selection for different diffusion amounts as a function of pressure.}
\end{figure}

The acceptance factor for each background and across different pressures for an enriched detector is shown in Figure~\ref{fig:EnrSelPerfPress}, considering the diffusions for 10\% He, 5\% CO$_2$, and no diffusion. Background rejection ranges from 90--99.9\% depending on the background category. 

Of the remaining backgrounds, the most frequent cases consist of topologies where a high energy delta electron is produced near the track start or gamma ray interactions produced from bremsstrahlung or Compton scatters deposit energy near the track start. Both of these signatures produce a stopping electron near the track that can mimic a two-electron $0\nu\beta\beta$ event signature. The most difficult backgrounds to reduce are the $^{137}$Xe, while the best background rejection is obtained for $^{208}$Tl events. This difference in performance is due to the subtle differences in topologies for how these backgrounds deposit energy in the ROI. $^{208}$Tl tends to have more multi-track topologies due to multiple gamma interactions while selected $^{137}$Xe background topologies have a higher fraction of topologies containing a high energy delta ray occurring near the track start. $^{214}$Bi sits between these two backgrounds as its photopeak is similar to producing a single electron in the ROI, like $^{137}$Xe backgrounds, while it can also deposit energy in the ROI through multiple gamma interactions, like $^{208}$Tl.

For the 10\% He sample, performance decreases with increasing pressure for the $^{137}$Xe background while $^{208}$Tl and $^{214}$Bi performance improves with pressure. This difference in background rejection with pressure is due to  deterioration in track reconstruction of the primary track with higher pressures, affecting variables that are harder to reconstruct, such as the blob energy. The $^{137}$Xe background is more affected by this degraded performance as these backgrounds rely more on the blob information. On the other hand, gamma backgrounds can be rejected more efficiently using more handles such as the total primary track energy and single track vs multi track topologies.

Similar to the 1 bar case, the overall performance improves with reducing the diffusion amount across higher pressures. In the case of the 5\% CO$_2$ and no diffusion all backgrounds show the same behavior of improved performance with pressure, due to the better quality of track reconstruction.

\section{Results}\label{sec:Enriched}

This section presents the overall acceptance factor and rates. We start with illustrating how the acceptance factors and rates change, first from events depositing energy in a 2.3 -- 2.6 MeV energy window, then by applying a cut assuming a fixed energy resolution of 0.75\% FWHM. The energy resolution cut applied is the same as described in Sec.~\ref{sec:Reconstruction}, which introduces an 18\% loss in signal efficiency. The 0.75\% FWHM energy resolution was chosen to show the acceptance and rate dependencies for one given energy resolution value. Finally, topological selection is applied assuming an enriched detector scanning three diffusion amounts across each pressure. The total rates are then also shown as a function of all energy resolutions and for different assumed detector technologies. 
\subsection{Acceptance and Rate with Diffusion and Pressure}

\begin{figure}[hbt]
\centering
\includegraphics[width=\textwidth]{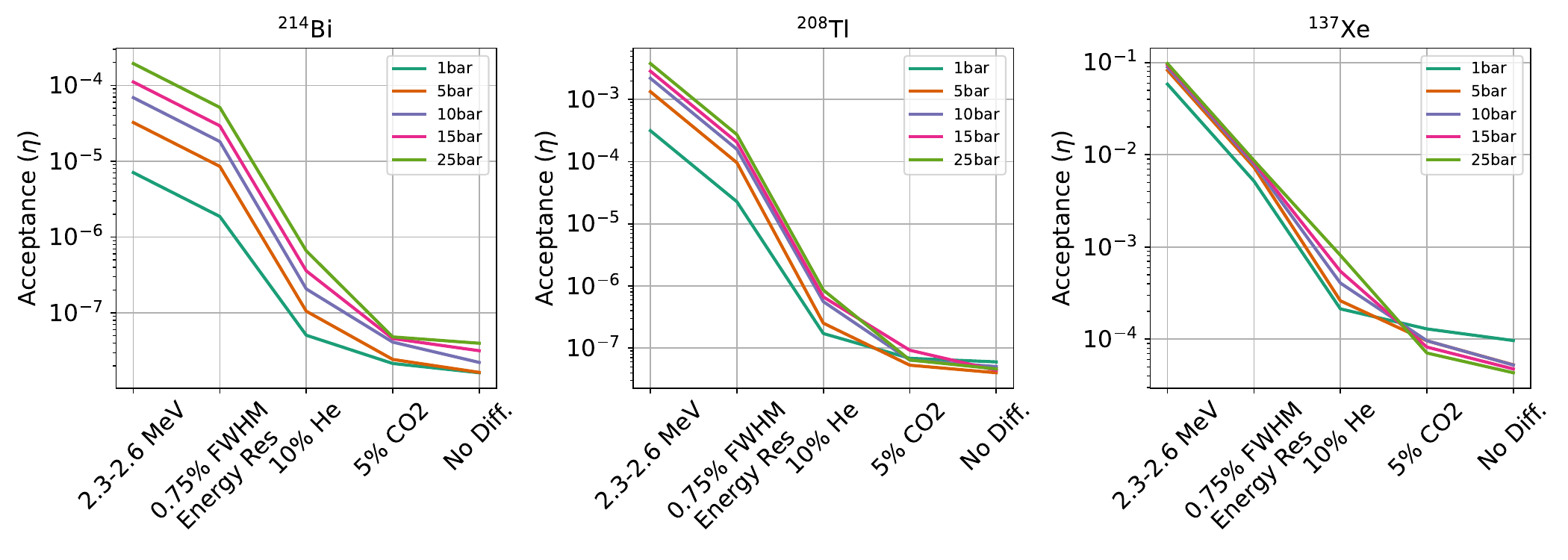}
\caption{\label{fig:EnrAcceptPress} The total acceptance factors for each background category as a function of the analysis cuts from left-to-right for an enriched detector configuration. The differences in acceptance, from 10\% helium to no diffusion, show the effect of reducing the diffusion on performance.   }
\end{figure}

The total acceptance factor for an enriched detector as a function of the analysis stage for each pressure and background category is shown in Fig.~\ref{fig:EnrAcceptPress}. The performance of a natural detector can be found in Appendix~\ref{sec:Natural}. Overall acceptance factors range from 10$^{-7}$--10$^{-8}$ for $^{208}$Tl and $^{214}$Bi while $^{237}$Xe acceptance factor is around 10$^{-4}$. 

The acceptance factors can be used to estimate the overall background rate per tonne-year:

\begin{equation}
    \textnormal{Rate} = \frac{\eta \cdot \mathrm{A} \cdot 3.15\times10^7}{\mathrm{M}_{\mathrm{Xe}}},
\end{equation}
where $\eta$ is the total acceptance factor, $\mathrm{A}$ is the total radioactivity in Bq, $3.15\times10^7$ is the number of seconds in a year, and $\mathrm{M}_{\mathrm{Xe}}$ is the mass of $^{136}$Xe in tonnes. The activity factors in the total mass of copper for $^{214}$Bi and $^{208}$Tl or total mass of $^{136}$Xe for $^{137}$Xe. 

For this work, we consider an activity of $^{238}$U and $^{232}$Th of 1.28 and 1.22~\textmu Bq/kg, respectively. The $^{137}$Xe assumes a site location at SNOLAB ($\sim$6000 m water equivalent (w.e.) depth) and a water shield surrounding the detector to reduce the neutron flux, amounting to a total activity of $\sim$1 count per year. A site location at LNGS ($\sim$3400 m w.e.) would increase the rate by a factor 100, and at LSC (2200 m w.e.) by a factor 1000~\cite{Rogers2020_Xe137Backgrounds}. 

\begin{figure}[hbt]
\centering
\includegraphics[width=\textwidth]{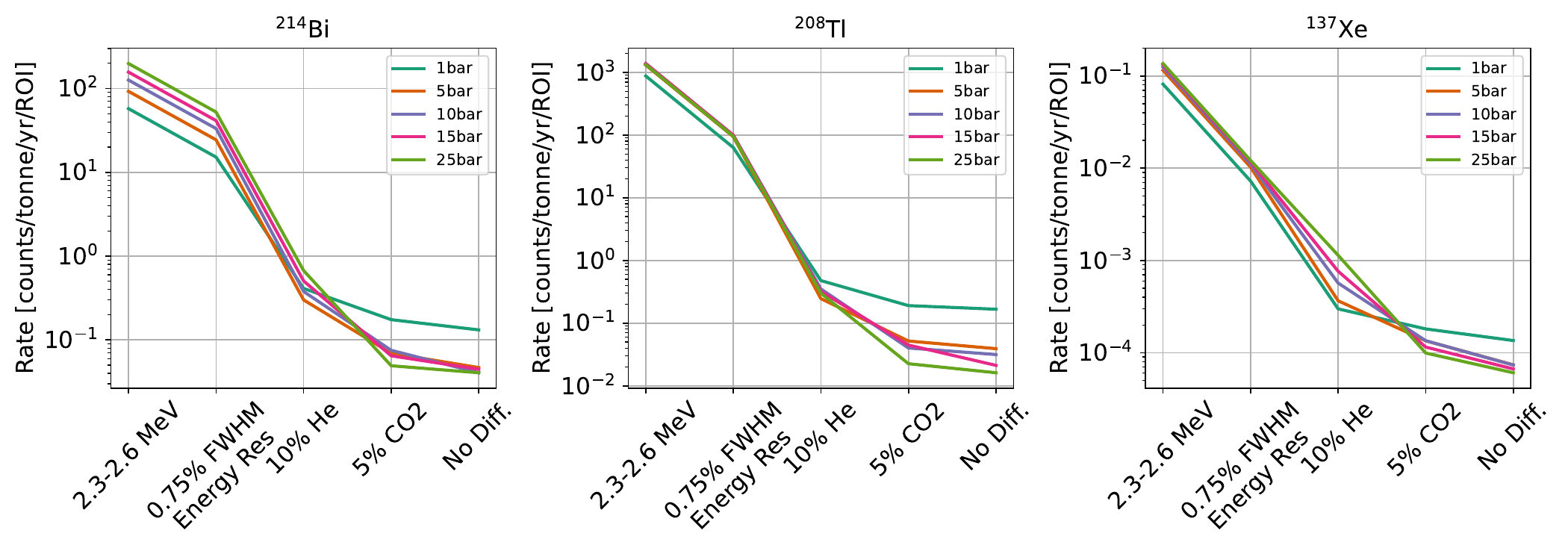}
\caption{\label{fig:EnrPerfPress} The total background rate for each background category as a function of the analysis cuts from left-to-right for an enriched detector configuration. The rates from 10\% helium to no diffusion show the effect of reducing the diffusion on performance.   }
\end{figure}

The rates for an enriched detector as a function of analysis stage are shown in Fig.~\ref{fig:EnrPerfPress}. The overall rates are below 0.1 counts/tonne/year/ROI for all pressures and backgrounds except 1 bar for $^{214}$Bi and $^{208}$Tl. The lower performance at 1 bar is due to the increased mass of copper and lower signal containment efficiency. The $^{137}$Xe rate follows the same trend as the acceptance, as the total intrinsic activity depends on the fixed isotopic mass of $^{136}$Xe.

\subsection{EL TPC}

\begin{figure}[hbt]
\centering
\includegraphics[width=\textwidth]{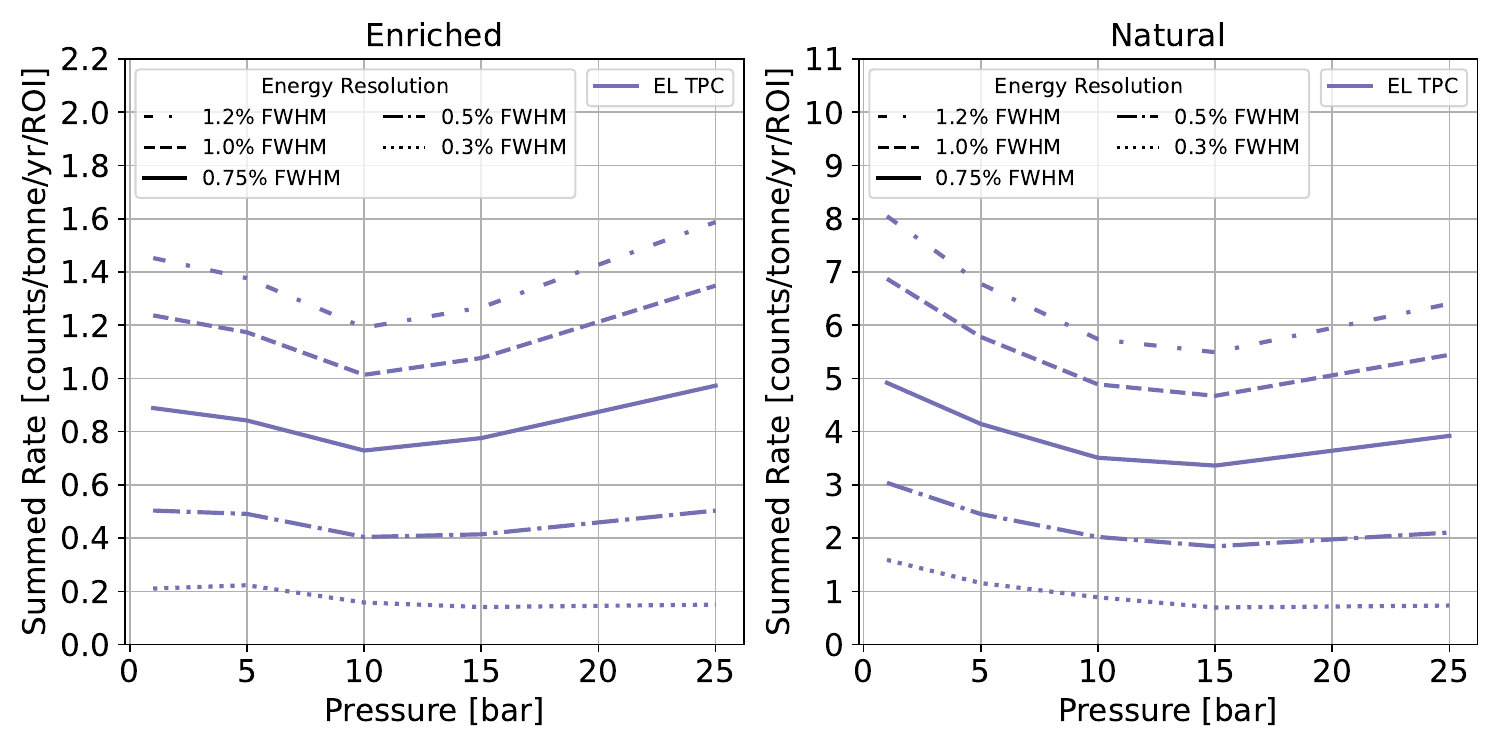}
\caption{\label{fig:PerfSummaryHe} A summary of the summed background rate for an EL TPC including $^{214}$Bi, $^{208}$Tl, and $^{136}$Xe. The left figure shows the performance for an enriched detector configuration, while the right figure shows a natural detector configuration.  }
\end{figure}

A summary of the summed background rates for an EL TPC assuming 10\% He gas admixture is shown in Fig.~\ref{fig:PerfSummaryHe}. These rates are summed over $^{214}$Bi, $^{208}$Tl, and $^{137}$Xe and consider an enriched and natural detector scanning the different assumed energy resolutions and pressures. Overall, the rates with pressure are mostly flat with a more optimal value around 10 bar pressure.

For the enriched EL TPC, background rates are below 1 count/tonne/year/ROI for energy resolutions below 0.75\% FWHM. Scanning across these varied energy resolutions can change the total rate by up to a factor of eight. A 0.5\% FWHM energy resolution is feasible with this technology yielding a total rate of 0.5 count/tonne/year/ROI for a one-tonne module. 

For the natural detector configuration, background rates are about a factor of 5 larger than those of the enriched detector. Rates similarly span a factor of eight across each energy resolution assumed.

\subsection{Topology TPC}
\begin{figure}[hbt]
\centering
\includegraphics[width=\textwidth]{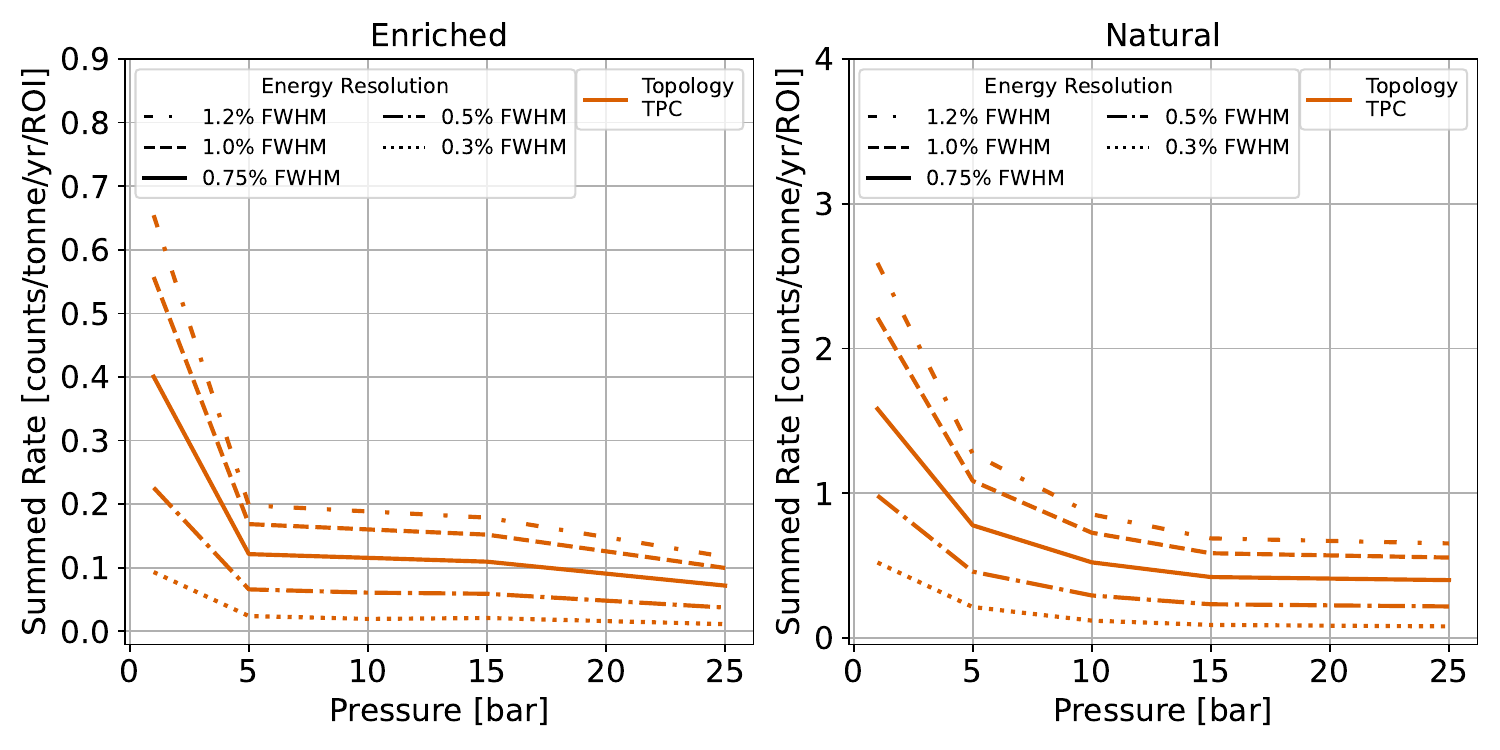}
\caption{\label{fig:PerfSummaryCO2} A summary of the summed background rate including $^{214}$Bi, $^{208}$Tl, and $^{136}$Xe, for a Topology TPC modeled with 5\% CO$_2$ gas additive. The left figure shows the performance for an enriched detector configuration, while the right figure shows a natural detector configuration. }
\end{figure}

In the case of a Topology TPC, we assume a 5\% CO$_2$ gas admixture to model the reduction in diffusion with a molecular gas additive. A summary of the summed background rates in this detector configuration, summed over $^{214}$Bi, $^{208}$Tl, and $^{137}$Xe, is shown in Fig.~\ref{fig:PerfSummaryCO2}. These rates are for an enriched and natural detector scanning the different assumed energy resolutions and pressures. 

Overall, the rates decrease with increasing pressure. Background levels just below 0.2 count/tonne/year/ROI are achieved for pressures greater than 5 bar. Similar to the EL TPC, the natural detector has a larger rate, reaching values below 1 count/tonne/year/ROI for pressures greater than 5 bar. 

As commented on in Sec.~\ref{sec:DetTech}, these performances assume the optimistic scenario where resolutions of at least 1.2\% FWHM are achieved and a fiducial volume cut can be applied with similar power to an EL TPC. We investigate the impact of removing the fiducial volume cut on these rates and radon backgrounds in Appendix~\ref{appdx:fv}. For the backgrounds originating from the detector surface, this introduces an increase in rate of about 0.1 counts/tonne/year/ROI, depending on the gas pressure and mixture. Background rates introduced from radon without suitable event placement would need to employ more advanced analysis techniques or explore new avenues for event vertex placement without VUV light to sufficiently tag the Bi-Po backgrounds.

\subsection{Ion TPC}
\begin{figure}[hbt]
\centering
\includegraphics[width=\textwidth]{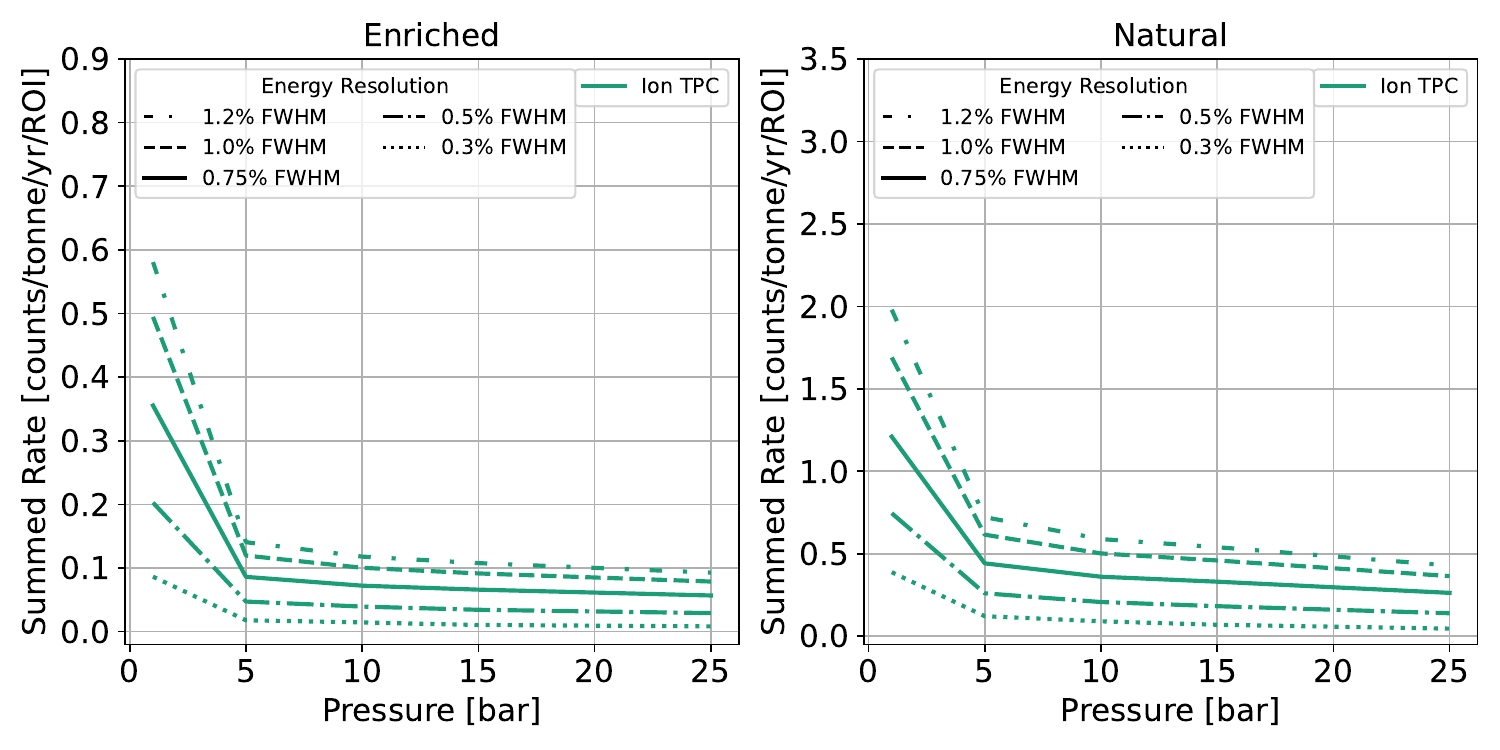}
\caption{\label{fig:PerfSummaryNoDiff} A summary of the summed background rate including $^{214}$Bi, $^{208}$Tl, and $^{136}$Xe, for an Ion TPC. The left figure shows the performance for an enriched detector configuration, while the right figure shows a natural detector configuration. }
\end{figure}

In the case of an Ion TPC, we assume no diffusion. A summary of the summed background rates in this detector configuration (over $^{214}$Bi, $^{208}$Tl, and $^{137}$Xe) is shown in Fig.~\ref{fig:PerfSummaryNoDiff}. These rates are for an enriched and natural detector scanning the different assumed energy resolutions and pressures.

The dependence and overall rates are similar to the Topology TPC. Likewise, these performances assume that such energy resolutions and suitable event placement are achieved.

\section{Discussion and Summary}\label{sec:conclusions}

One of the major limitations of the natural detector configuration is the large mass of copper required to build a detector of that size. The possibility of operating the natural detector with 5--10 tonnes of $^{136}$Xe could help boost the signal-to-background ratio to yield a similar performance to the enriched scenario. In this case, the $^{137}$Xe backgrounds would also increase proportionally to the $^{136}$Xe mass; however, this could be tolerated at a site such as SNOLAB, resulting in a rate below a fraction of a count/tonne/year/ROI. Similarly, the 1 bar suffers from the same issue of large copper mass and low signal containment efficiency. New engineering solutions to mitigate the copper background, and/or more effective analysis selections utilizing the higher track clarity are well motivated for this detector configuration. 

For pressures above 5 bar, the total background rates are flat. A further option would be to operate an enriched detector at 5 bar and then load in more isotope with increased pressure to pack in more source isotope. Detector design should optimize sensor readout and vessel design for the highest pressure planned to ensure suitable track clarity is maintained. 

This work also considers a simple analysis for topological rejection in order to scan many parameters. Machine learning methods, which can account for the full topology of the track rather than using the end-point information, have been shown to further improve performance with respect to a simple analysis~\cite{JINST_T01004_2017,Kekic2021_NEXTCNN} with potential for also improving signal efficiency. Furthermore, post-processing methods to reduce the effect of track blurring due to diffusion, such as deconvolution, have shown significantly improved background rejection performance~\cite{NEXT2021_RLDeconvolution}. The largest scope for improvement with more advanced reconstruction methods is at lower pressures, which can utilize the higher track detail to better reject backgrounds, such as identifying energetic delta rays occurring near the track start, which have a similar signature of two electrons emitted from a common vertex, or tagging of atomic x-rays (K,L-shell) from photoelectric conversions from gamma-induced backgrounds. 

In summary, this work investigates how the possible design choices can impact a gaseous xenon TPC searching for $0\nu\beta\beta$ towards a half-life sensitivity of 10$^{27}$--10$^{28}$ years considering a detector with one tonne of source isotope. We consider three of the major sources of background for the experiment arising from $^{137}$Xe, $^{214}$Bi, and $^{208}$Tl radioactive decays. Several variables are surveyed, including detector size accounting for gas enrichment, copper shielding mass, gas pressure, energy resolution, and analysis performance. A detector configuration optimized for accommodating xenon enriched with $^{136}$Xe at 90\% is preferred over operation with natural xenon with an order of magnitude lower background rate. This is due to the increase in intrinsic background due to the larger detector size and inner copper shielding mass, which overcomes the improvement in signal containment efficiency and analysis-level rejection power.

The performance of three gas TPC technologies was also explored based on different gas additives used to reduce diffusion. An EL TPC includes 10\% helium preserving electroluminescence signals based on the well-established NEXT experiment detector design. The total rate is expected to be below 0.5 counts/tonne/year/ROI, assuming an energy resolution of 0.5\% FWHM. The Topology and Ion TPCs include the addition of stronger diffusion-reducing molecular additives and could produce background rates below 0.2 counts/tonne/year/ROI. Such performances are contingent on an accurate 3-D event placement method and energy resolution of 1.2\% FWHM or less, achieved without the use of VUV light signals. 

In addition to dependence on site location, enrichment level and analysis level choices, the performance of a tonne-scale GXeTPC is shown to be dependent on gas pressure. Construction challenges, beyond the scope of this paper, are substantial at any pressure, balancing different considerations such as high voltage requirements or detector size. Altogether, at this level of investigation there is no clearly optimum pressure choice, yet further study involving machine learning may reveal important advantages. We anticipate a follow-on paper with explicit design choices for a tonne-scale GXeTPC experiment.

\acknowledgments
{ The authors would like to thank Ben Jones, Adam Para, Juan José Gomez Cadenas, Pau Novella, and Michel Sorel for their valuable feedback on this manuscript. This research used services provided by the OSG Consortium \cite{osg1,osg2,osg3,osg4}, which is supported by the National Science Foundation awards \#2030508 and \#1836650. We also acknowledge support from the following agencies: DE-SC0019054 (Texas Arlington) and DE-SC0019223 (Texas Arlington).}

\bibliographystyle{JHEP}
\bibliography{biblio}


\appendix

\section{Natural Xenon Detector}\label{sec:Natural}

\begin{figure}[hbt]
\centering
\includegraphics[width=\textwidth]{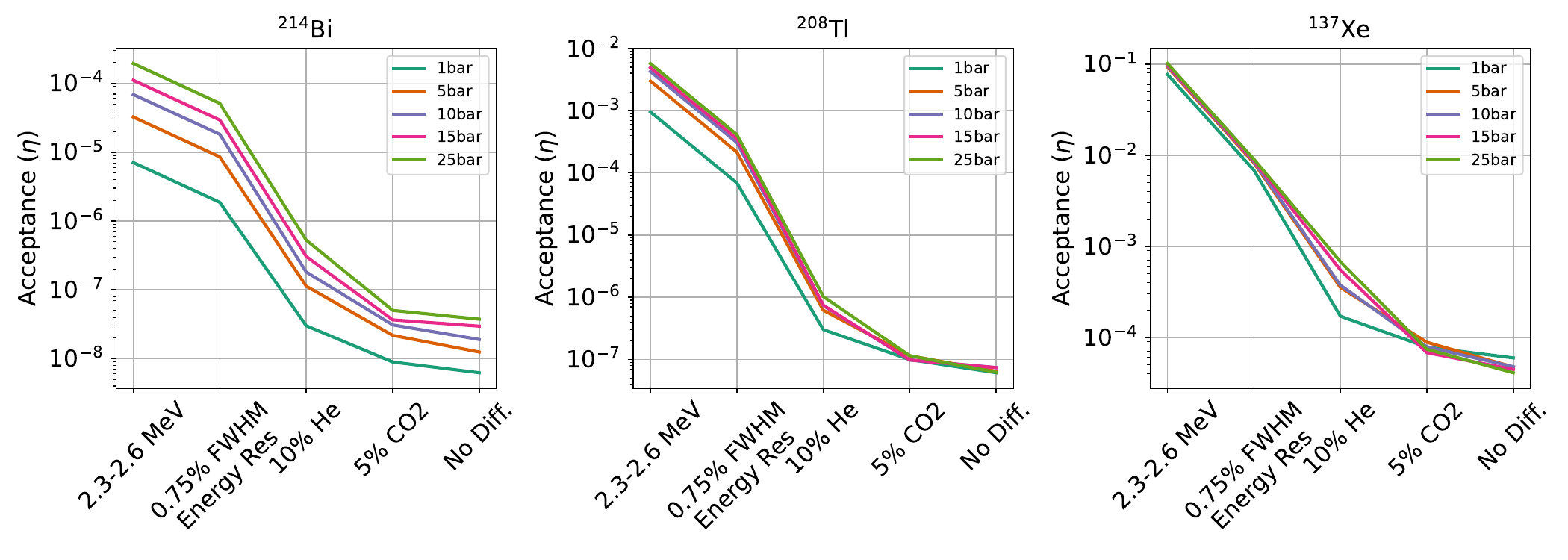}
\caption{\label{fig:NatAcceptPress} The total acceptance factors for each background category as a function of the analysis cuts from left-to-right for a natural detector. The acceptance from 10\% helium to no diffusion shows the effect of reducing the diffusion on performance.  }
\end{figure}

The acceptance factors for a natural detector are shown in Fig.~\ref{fig:NatAcceptPress}. Overall, the acceptance factors slightly improve on the enriched detector due to the higher signal containment. 

\begin{figure}[hbt]
\centering
\includegraphics[width=\textwidth]{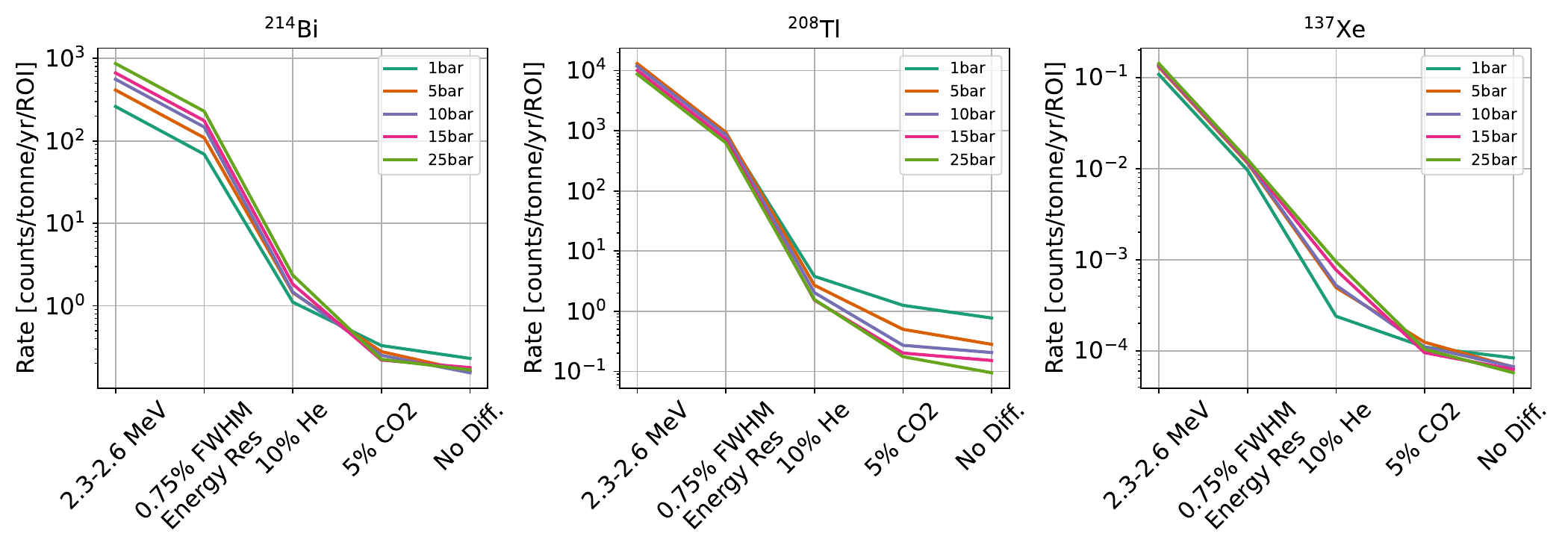}
\caption{\label{fig:NatPerfPress} The total background rate for each background category as a function of the analysis cuts from left-to-right for a natural detector. The rates from 10\% helium to no diffusion shows the effect of reducing the diffusion on performance.    }
\end{figure}

The total background rate for the natural detector is shown in Fig.~\ref{fig:NatPerfPress}. While there is a similar background rate for $^{137}$Xe, the $^{214}$Bi and $^{208}$Tl backgrounds do not perform as well due to the large mass copper shielding required. 

\section{Impact of Event Placement}\label{appdx:fv}

The fiducial volume cut is powerful in reducing the surface background, such as Bi-Po from the copper surface and cathode. For technologies such as the Topology and Ion TPCs, S1 light may be quenched through the addition of molecular additives, requiring alternative ways to be developed to place events in the drift direction suitably. We explore the effectiveness of using topology selections without event position placement for (i) surface backgrounds and (ii) radon-induced backgrounds. 

\begin{figure}[hbt]
\centering
\includegraphics[width=\textwidth]{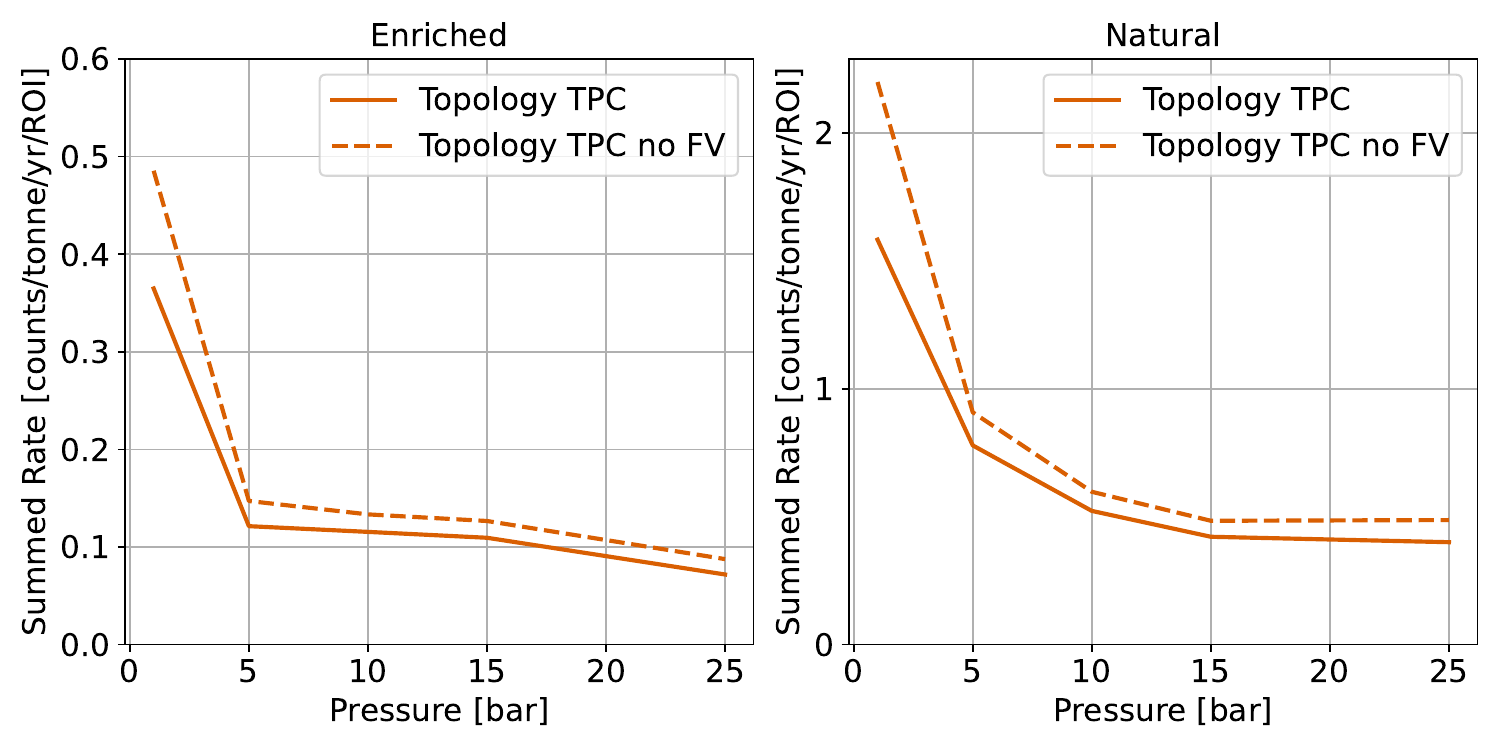}
\caption{\label{fig:PerfSummaryFV} A summary of the summed background rate with and without the fiducial volume (FV) cut including $^{214}$Bi, $^{208}$Tl, and $^{136}$Xe, for a Topology TPC assuming 5\% CO$_2$ gas additive. The left figure shows the performance for an enriched detector configuration, while the right figure shows a natural detector configuration. These performances assume a 0.75\% FWHM energy resolution.}
\end{figure}

\textbf{Surface Background:} Figure~\ref{fig:PerfSummaryFV} shows the rate with and without the application of the fiducial volume cut of 2~cm applied to the wall edges in a Topology TPC (assuming 5\% CO$_2$ gas additive). For the enriched detector, the rate increase for 5\% CO$_2$ is slightly larger at lower pressure with a maximal rate increase of $\sim$0.1 counts/tonne/yr/ROI. The natural detector also follows a similar trend with an increase in rate at lower pressure compared with higher. 

\begin{figure}[hbt]
\centering
\includegraphics[width=0.45\textwidth]{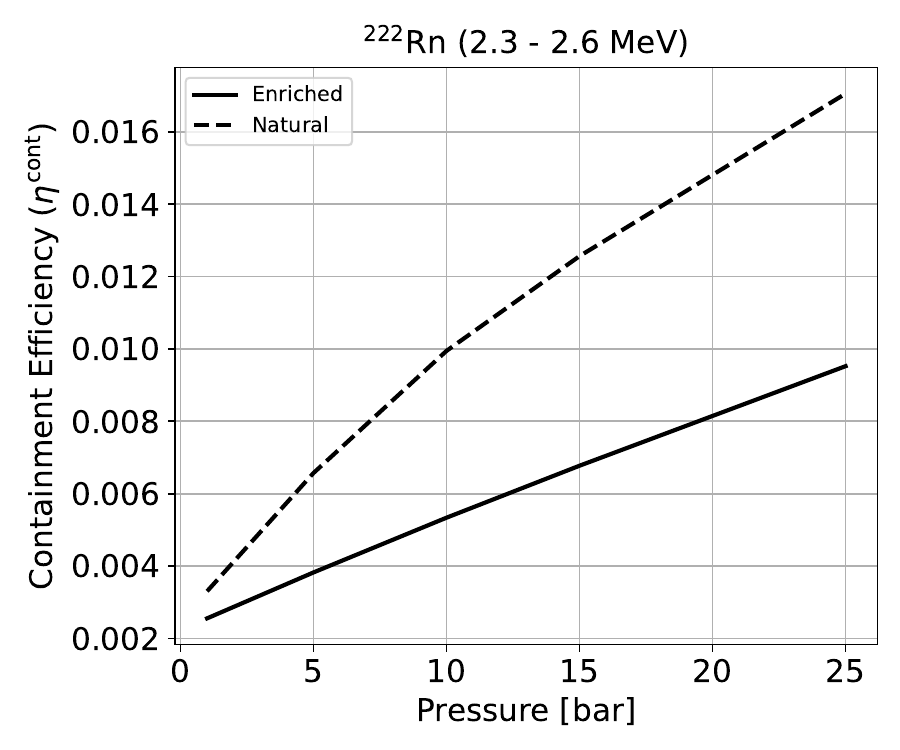}
\includegraphics[width=0.45\textwidth]{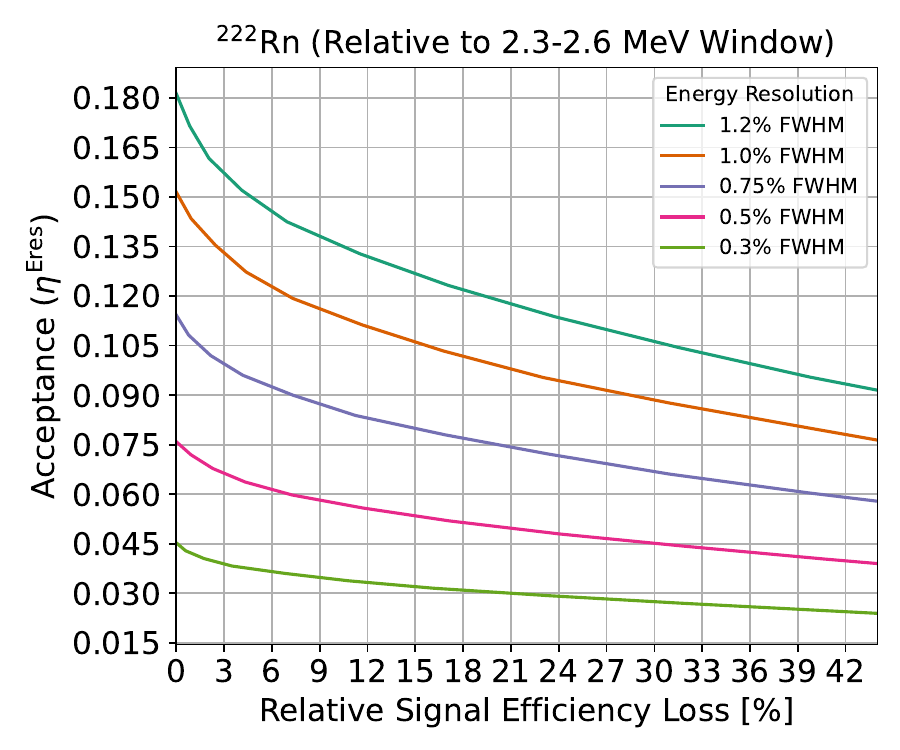}
\caption{\label{fig:RadonContEres} (left) The containment rates for $^{214}$Bi at the cathode induced from $^{222}$Rn as a function of pressure for an enriched and natural detector. (right) The background acceptance rates assuming different energy resolutions for $^{214}$Bi at the cathode induced from $^{222}$Rn. The x-axis shows the efficiency loss vs background acceptance from increasing the lower bound cut on energy. }
\end{figure}

\textbf{Radon:} To study the impact of $^{214}$Bi at the cathode induced from $^{222}$Rn, we follow the same simulation procedure as described in Sec.~\ref{sec:Detector}. $^{214}$Bi events are generated (including subsequent decays and deexcitations to $^{210}$Pb) at the surface of the detector face with the longest drift distance to model the placement of a cathode. $^{214}$Bi decays were simulated resulting in $\sim$200k events depositing energy within 1\% FWHM of the Q$_{\beta\beta}$-value for each pressure. The containment rates and energy resolution cuts for this background are shown in Fig.~\ref{fig:RadonContEres}. The containment fractions are increasing with pressure, while the reduction in background from energy resolution is similar in magnitude to $^{208}$Tl/$^{137}$Xe.

\begin{figure}[hbt]
\centering
\includegraphics[width=\textwidth]{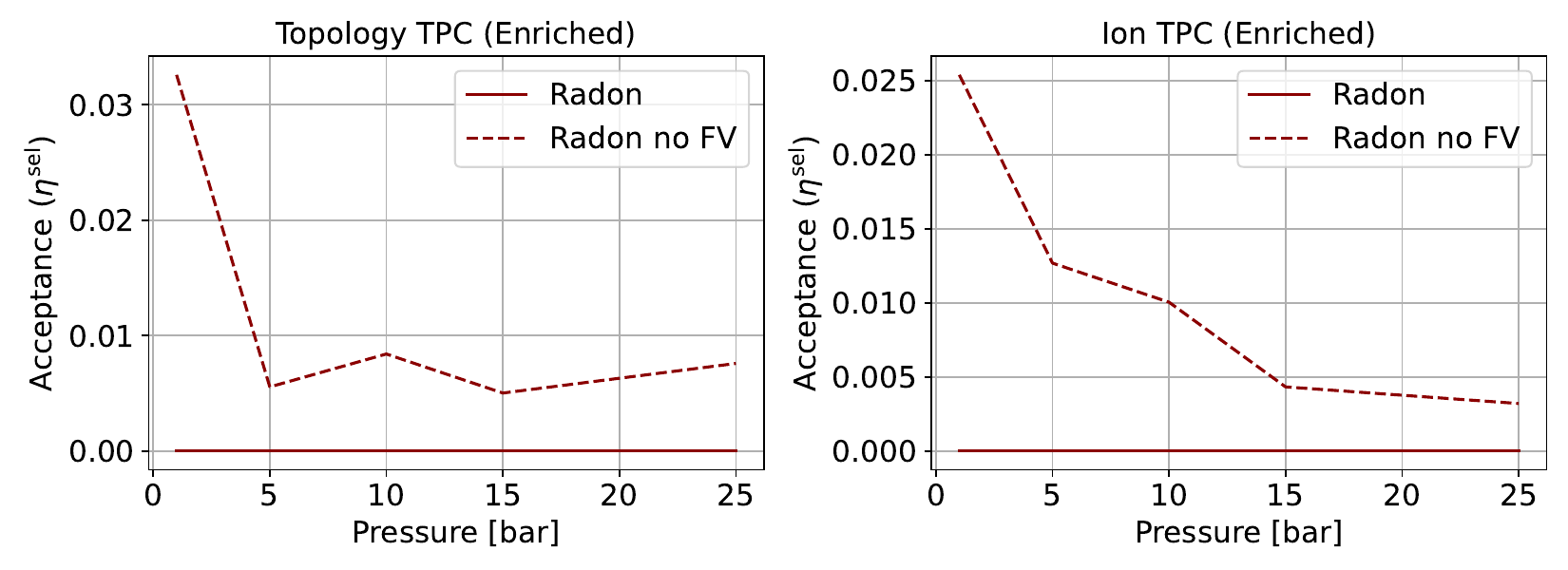}
\caption{\label{fig:PerfSummaryRadon} The selection rejection factors with and without a fiducial volume cut for (left) a Topology TPC and (right) an Ion TPC. A fiducial volume cut is able to reduce the radon background to negligible levels while some background remains without the fiducial volume cut. }
\end{figure}

The selection performance with and without using a fiducial volume cut is shown in Fig.~\ref{fig:PerfSummaryRadon}. Without any reconstruction inefficiencies, the radon background is reduced to a negligible level with a 2~cm fiducial volume cut. Removing the fiducial volume cut, background rejection factors are around 97-99.5\% depending on pressure. 

Radon activity reported by the NEXT-White experiment is 38.1 mBq/m$^3$~\cite{NEXT2018_Radon}. This background rate depends on the detector volume. The total acceptance factors using event topology without the fiducial volume cut range from 2-7$\times10^{-6}$ and would be insufficient for rejecting this background without additional means. 

Of the selected radon events, a majority are from the 3.2 MeV beta electron stretching into the gas region. This is combined with a large energy deposition at the track start due to partial energy deposition from the 7.8~MeV alpha from the decay of $^{214}$Po which reduces the effectiveness of the blob energy cut. Analyses may utilize more advanced topological features, such as the straightness of the beta track near the track start or the larger diffusion to tag these cathode-originating events and improve performance. Furthermore, alternative methods for suitable event placement could be employed with this technology, such as the identification of positive ion signals reaching the cathode. 

\end{document}